\documentclass[11pt]{article}
\usepackage{graphicx}
\usepackage{amsmath}
\InputIfFileExists{Debug}{}{}

\usepackage{amsmath}
\usepackage{amsthm}
\usepackage{amsfonts}          % if you want the fonts
\usepackage{amssymb}           % if you want extra symbols
\usepackage[mathscr]{eucal}
\usepackage{graphicx}
\usepackage{color}
\usepackage{hypbmsec}
\usepackage{fancyvrb}
\usepackage{sepnum}

\usepackage{rotating}
\usepackage{slashbox}
\usepackage[isu,bf]{caption}
\usepackage[all]{xy}           % Commutative diagrams

\newlength{\xtrawidth}
\newlength{\xtraheight}
\ifx\NoBackreferences\EMPTYMACRO
  \usepackage[backref,linktocpage,bookmarks]{hyperref}
\else
  \usepackage[linktocpage,bookmarks]{hyperref}
\fi

\ifx\DEBUG\EMPTYMACRO
\else
\fi

\def\clap#1{\hbox to 0pt{\hss#1\hss}}

\def\mathclap{\mathpalette\mathclapinternal}

\def\mathclapinternal#1#2{%
\clap{$\mathsurround=0pt#1{#2}$}}	

\makeatletter
  \def\adots{\mathinner{\mkern2mu\raise\p@\hbox{.}
      \mkern2mu\raise4\p@\hbox{.}\mkern1mu
      \raise7\p@\vbox{\kern7\p@\hbox{.}}\mkern1mu}}
\makeatother

\newcommand{\C}{\ensuremath{{\mathbb{C}}}}

\newcommand{\Z}{\mathbb{Z}}
\newcommand{\CP}{\ensuremath{\mathop{\null {\mathbb{P}}}}\nolimits}

\newcommand{\Ncal}{\mathcal{N}}

\DeclareMathOperator{\Span}{span}
\DeclareMathOperator{\Pic}{Pic}

\DeclareMathOperator{\tr}{tr}

\DeclareMathOperator{\rank}{rank}

\DeclareMathOperator{\End}{End}

\DeclareMathOperator{\Vol}{Vol}
\DeclareMathOperator{\dVol}{dVol}

\newcommand{\Lsheaf}{\ensuremath{\mathscr{L}}}
\newcommand{\Osheaf}{\ensuremath{\mathscr{O}}}

\newcommand{\Vsheaf}{\ensuremath{\mathscr{V}}}

\newcommand{\Kcone}{\ensuremath{\mathcal{K}}}

{\end{list}}

{\everymath{\displaystyle\everymath{}}\array}%
{\endarray}

\newtheorem{theorem}{Theorem}

\numberwithin{equation}{section}

\newcommand{\beq}{\begin{equation}}
\newcommand{\eeq}{\end{equation}}

\newcommand{\ba}{\begin{array}}
\newcommand{\ea}{\end{array}}

\newcommand{\rk}{\mathop{{\rm rk}}}

\newcommand{\cO}{{\cal O}}

\newcommand{\cF}{{\cal F}}

\newcommand{\cL}{{\cal L}}

\begin{document}

\begin{titlepage}
  \vspace*{-2cm}
  \hfill
 % \parbox[c]{5cm}{
    %\begin{flushright}
     % DIAS-STP [FIXME]
    %\end{flushright}
  %}
  \vspace*{2cm}
  \begin{center}
    \huge
    Numerical Hermitian Yang-Mills Connections  \\
    and K\"ahler Cone Substructure
  \end{center}
  \vspace*{8mm}
  \begin{center}
    \begin{minipage}{\textwidth}
      \begin{center}
        \sc 
        Lara B.\ Anderson${}^{1}$,
        Volker Braun${}^{2}$, and
        Burt A.\ Ovrut${}^{1}$
      \end{center}
      \begin{center}
        \textit{
          ${}^1$Department of Physics, University of Pennsylvania\hphantom{${}^1$}\\ 
          209 South 33rd Street, Philadelphia, PA 19104-6395, U.S.A.
        }\\[1ex]
        \textit{
          ${}^2$Dublin Institute for Advanced Studies\hphantom{${}^2$}\\
          10 Burlington Road, Dublin 4, Ireland.
        }\\[1ex]
        
      \end{center}
    \end{minipage}
  \end{center}
  \vspace*{\stretch1}
  \begin{abstract}
    We further develop the numerical algorithm for computing the gauge
    connection of slope-stable holomorphic vector bundles on
    Calabi-Yau manifolds.  In particular, recent work on the generalized Donaldson algorithm is 
    extended to bundles with K\"ahler
    cone substructure on manifolds with $h^{1,1}>1$. Since the computation depends only on a
    one-dimensional ray in the K\"ahler moduli space, it can probe
    slope-stability regardless of the size of $h^{1,1}$. Suitably
    normalized error measures are introduced to quantitatively compare
    results for different directions in K\"ahler moduli space. A
    significantly improved numerical integration procedure based on
    adaptive refinements is described and implemented. Finally, an
    efficient numerical check is proposed for determining whether or
    not a vector bundle is slope-stable without computing its full
    connection.
  \end{abstract}
  \vspace*{\stretch1}
  \begin{center}
    \texttt{Email: 
      andlara@physics.upenn.edu, 
      vbraun@stp.dias.ie,
      ovrut@elcapitan.hep.upenn.edu
    }
  \end{center}
\end{titlepage}

\tableofcontents

\section{Introduction}
In this paper, we explore $\Ncal=1$ supersymmetric vacua of $E_8
\times E_8$ heterotic string~\cite{Gross:1984dd,Candelas:1985en} and
$M$-theory~\cite{Horava:1995qa, Horava:1996ma, Witten:1996mz,
  Lukas:1997fg, Lukas:1998yy, Lukas:1998tt}. The four-dimensional
effective theory is specified by a Calabi-Yau threefold $X$ and a
slope-stable holomorphic vector bundle, $\Vsheaf$. The detailed
structure of the low energy theory is determined~\cite{Green:1987sp}
by the choice of a Ricci-flat metric, $g$, on the threefold and an
$\Ncal=1$ supersymmetry gauge connection, $A$, on the vector
bundle. Existence proofs, such as Yau's theorem~\cite{MR480350} for
the Ricci-flat metric and the Donaldson-Uhlenbeck-Yau
theorem~\cite{duy1,duy2} for the Hermitian Yang-Mills connection,
provide us with numerous examples of such geometries. However, the
explicit metric and gauge connection are not known analytically,
except in very special cases~\cite{Candelas:1987rx, Candelas:1990rm,
  Greene:1993vm, Donagi:2006yf, Braun:2006me,Anderson:2009ge}. The difficulty of
determining these quantities has presented an obstacle to the
systematic search for realistic heterotic vacua. Even for a known
vacuum, this has precluded the computation of physically relevant
parameters in the effective theory, such as the Yukawa couplings.

In recent years, the development of sophisticated numerical
approximation schemes have provided a new approach to these
problems~\cite{MR2161248,MR1916953, DonaldsonNumerical, MR1064867,
  MR2154820,Douglas:2006hz,
  MR2194329,Doran:2007zn,Headrick:2009jz,Douglas:2008es}. With the
development of powerful new algorithms and modern computer speed, it
is now possible to numerically approximate Ricci-flat metrics and
Hermitian Yang-Mills connections to a high degree of accuracy. We will
refer to these tools collectively as the ``generalized Donaldson
algorithm". Using them, the structure of the four-dimensional
effective theory can be explored in remarkable new ways. The goal of
this program is to determine all coefficients in the superpotential,
the explicit form of the K\"ahler potential and, ultimately, to
perform first-principle calculations of physical quantities such as
the relative quark and lepton masses.

In this paper, we make substantial progress towards this goal by
extending previous work~\cite{Anderson:2010ke} to include vector
bundles defined over manifolds with higher-dimensional K\"ahler cones;
that is, for which $h^{1,1}>1$. Importantly, our results allow one to
study arbitrary vector bundles arising in heterotic string
compactifications and to determine whether such geometries admit
$\Ncal=1$ supersymmetric vacua. The problem of finding the
K\"ahler cone substructure, that is, the regions in K\"ahler moduli
space where a given holomorphic vector bundle is or is not
slope-stable, is a notoriously difficult one. In particular, the
difficulty of a direct stability analysis generally increases rapidly
with the dimension $h^{1,1}$ of the K\"ahler cone. One of the great
advantages of the algorithm presented in this paper is that, unlike a
standard analytic analysis, our numerical calculations can be
performed with essentially equal ease in arbitrary $h^{1,1}$. This
provides an important new tool in the study of supersymmetric
heterotic vacua. 

The structure of the paper is as follows. To begin, in
\autoref{review_section} we provide a brief review of the numerical
algorithm for computing the Ricci-flat metric, $g$, and the Hermitian
Yang-Mills connection, $A$. Starting in \autoref{metric_review}, an
overview of Donaldson's algorithm for computing the Ricci flat
metric~\cite{MR2161248, MR1916953, DonaldsonNumerical} on a Calabi-Yau
manifold is given. In particular, the numerical implementations
developed in~\cite{Braun:2007sn, Braun:2008jp}
and~\cite{Douglas:2006rr, Douglas:2006hz} are discussed.  Next, in
\autoref{connection_alg}, we outline the recent generalizations of
Donaldson's algorithm presented
in~\cite{MR2154820, Douglas:2006hz, Anderson:2010ke}. These make it
possible to compute Hermite-Einstein metrics on holomorphic vector
bundles over a Calabi-Yau manifold and, hence, to solve for the unique
gauge connection satisfying the conditions for $\Ncal=1$
supersymmetry.

Both the original Donaldson algorithm and its generalizations to
connections rest on finding a particularly ``nice'' projective
embedding. In the
case of the Ricci-flat metric on a Calabi-Yau manifold, the embedding
is defined from $X$ into some higher-dimensional projective space via
the global sections of some ample line bundle, $\cL^{\otimes k_g}$, on $X$. In
the case of the connection on a rank $n$ bundle $\Vsheaf$, a map into
the Grassmannian $G(n,N_{k_{H}}-1)$ is constructed out of the
$N_{k_{H}}$ sections of $\Vsheaf \otimes \cL^{\otimes k_{H}}$, where $
\cL^{\otimes k_{H}}$ is some ample line bundle on $X$. Using either
one of these embeddings\footnote{Technically, we only require a map that is an immersion rather than an embedding. However, we need not make the distinction in this paper.}, a metric can be pulled-back to the Calabi-Yau
manifold and vector bundle, respectively. One obtains a
$k_g$-dependent sequence of K\"ahler metrics $X$ and a $k_H$-dependent
sequence of Hermitian fiber metrics on $\Vsheaf$. The degrees of
freedom of every embedding parametrize a family of pulled-back
metrics. By tuning the embedding to the so-called ``balanced embedding'' for
each degree $k_g$ and $k_H$, the K\"ahler metrics converge to the Ricci-flat
metric on $X$ and the Hermitian fiber metrics converge to a
Hermite-Einstein metric on $\Vsheaf$. Finding the balanced embedding
is solved by Donaldson's T-operator and its generalization due to Wang
and others~\cite{MR2161248, 2008arXiv0804.4005S, MR1916953,
  DonaldsonNumerical, MR1064867, MR2154820, 2007arXiv0709.1490K,
  2002math......3254P}. Roughly, the T-operator acts on embeddings of
fixed degree and has the balanced embedding as a fixed point. For a
Calabi-Yau manifold, iterating the T-operator will always converge to
a balanced embedding, and, therefore, to the Ricci-flat metric on $X$
in the limit that $k_g \to \infty$. In the case of the connection, the
iteration of the T-operator for fixed $k_H$ is not guaranteed to
converge. In fact, it converges to a balanced embedding if and only if
the bundle $\Vsheaf$ is Gieseker-stable. Furthermore, when $\Vsheaf$
is slope-stable then the sequence of balanced embeddings define a
fiber metric converging to the Hermite-Einstein fiber metric in the
limit that $k_H \to \infty$.

In \autoref{substruc_sec}, we modify a number of the numerical tools
developed in previous work~\cite{Anderson:2010ke} to enable us to
compare the convergence of the generalized Donaldson algorithm for
different rays (or ``polarizations") in K\"ahler moduli space. These
results are illustrated in \autoref{eg_sec} by an indecomposable, rank
$2$ vector bundle defined over the $K3$ surface via the monad
construction~\cite{okonek, Anderson:2007nc,
  Anderson:2008uw}. Furthermore, in \autoref{volker_int}, we present
the technical details of a newly developed, rapid numerical scheme for
integrating over a Calabi-Yau manifold.

\autoref{t_iter_error} provides a criterion to decide whether or not a
vector bundle is slope-stable for a given polarization, without the
need to explicitly compute the connection.  Hence, this check can be
rapidly applied to decide slope stability. In particular, we use a
result of Wang~\cite{MR2154820} which states that the iteration of the
T-operator will reach a fixed point if and only if the defining vector
bundle is Gieseker-stable. While Gieseker stability is not sufficient
to guarantee a solution to the Hermitian Yang-Mills equations (and,
hence, a supersymmetric heterotic vacuum), it still provides valuable
information. In particular, while a slope-stable bundle is
automatically Gieseker stable, the converse does not follow. A
Gieseker-stable bundle need only be slope semi-stable. Despite these
subtleties, we extract results from the T-operator convergence which
can be used to determine the K\"ahler cone substructure. Related to
the question of semi-stability, we consider the dependence of
slope-stability on the vector bundle moduli $H^1(\Vsheaf \otimes
\Vsheaf^{\vee})$ in \autoref{stab_wall_sec}. In particular, the
numerical algorithm is tested on a ``stability
wall''~\cite{Anderson:2009sw,Anderson:2009nt} in K\"ahler moduli
space, the boundary between slope-stable and unstable regions. We find
that the generalized Donaldson algorithm is sensitive to the bundle
moduli dependence and, hence, our results also distinguish the
marginal cases of slope poly-stable bundles from strictly semi-stable
ones.

In \autoref{cy3_sec}, we extend our study to higher-dimensional spaces by
presenting an example of K\"ahler cone substructure of a rank $3$
monad bundle defined over a Calabi-Yau threefold constructed as a
complete intersection in a product of projective spaces. We conclude
and discuss future work in \autoref{conclusion_sec}.

\section{Hermitian Yang-Mills Connections and Fiber Metrics}
\label{review_section}

A supersymmetric $E_8 \times E_8$ heterotic string compactification is
specified by 1) a complex $d$-dimensional Calabi-Yau manifold, $X$,
and 2) a holomorphic vector bundle, $\Vsheaf$, with structure group $K
\subset E_{8}$ defined over $X$. The gauge connection, $A$, on
$\Vsheaf$ with associated field strength, $F$, must satisfy the
well-known Hermitian Yang-Mills (HYM)
equations~\cite{Green:1987sp}. For general $U(n)$ structure groups,
these equations are given by
\begin{equation}
  \label{hym_genr}
  F_{ij}=F_{\bar{i}\bar{j}}=0
  ,\qquad
  g^{i\bar{j}}F_{i\bar{j}}
  =
  \mu(\Vsheaf)\cdot \mathbf{1}_{n \times n} , 
\end{equation} 
where $ g^{i\bar{j}}$ is the Calabi-Yau metric, $n$ is the rank of
$\Vsheaf$, the scalar $\mu(\Vsheaf)$ is a real number associated with
$\Vsheaf$ and $i,j=1,\ldots d$, run over the holomorphic indices of
the Calabi-Yau $d$-fold. Our primary interest is in Calabi-Yau
threefolds, since compactification on these give rise to $\Ncal=1$
supersymmetric theories in four dimensions. However, in order to
present simple illustrations of the techniques introduced in this
paper, we will discuss Calabi-Yau twofold ($K3$) as well as threefold
examples. It is not strictly necessary for the first Chern class of
the bundle to vanish~\cite{Blumenhagen:2006ux}, and the methods used
in this paper would work just as well in that setting. However, most
realistic compactifications are based on structure groups $K=SU(n)
\subset U(n)$, and these will be our main focus. When $K=SU(n)$, the
parameter $\mu(\Vsheaf)=0$ and eq.~\eqref{hym_genr} reduces to
\begin{equation}
  \label{the_hym}
  F_{ij}=F_{\bar{i}\bar{j}}=0
  ,\qquad
  g^{i\bar{j}}F_{i\bar{j}}=0 .
\end{equation} 
While eq.~\eqref{the_hym} are the relevant equations for realistic
heterotic compactifications, mathematically it will often be useful to
discuss the Hermitian Yang-Mills equations in full generality.
%For $K=SU(n)$, the structure groups of interest in this paper, these become
%\begin{equation}
%  \label{the_hym}
%  F_{ij}=F_{\bar{i}\bar{j}}=0
%  ,\quad
%  g^{i\bar{j}}F_{i\bar{j}}=0
%\end{equation} 
%where $ g^{i\bar{j}}$ is the Calabi-Yau metric and $i,j=1,\ldots d$, run over the holomorphic indices of the Calabi-Yau
%$d$-fold. For realistic physics, our primary interest rests on Calabi-Yau threefolds which can give rise to $\mathcal{N}=1$ supersymmetric theories in $4$-dimensions, however as in illustration of the techniques in this work, we will also consider Calabi-Yau two-folds (and vector bundles over them) as examples. Furthermore, equations \eqref{the_hym} are a special case of the more general $U(n)$ Hermitian Yang-Mills
%equations 
%\begin{equation}
%  \label{hym_genr}
%  g^{i\bar{j}}F_{i\bar{j}}
 % =
%  \mu(\Vsheaf)\cdot \mathbf{1}_{n \times n} , 
%\end{equation} 
%where $n$ is the rank of $\Vsheaf$ and $\mu(\Vsheaf)$ is a quasi-topological real number associated to $\Vsheaf$. When $K=SU(n)$, $\mu(\Vsheaf)=0$ and, \eqref{hym_genr} reduces to \eqref{the_hym}. While, \eqref{the_hym} are the relevant equations for a heterotic compactification, mathematically it will be useful in this work to discuss the Hermitian Yang-Mills equations in full generality.

A solution to \eqref{hym_genr} is equivalent to the
bundle $\Vsheaf$ carrying a particular Hermitian structure. An 
Hermitian structure (or Hermitian fiber metric), $G$, on $\Vsheaf$ is
an Hermitian scalar product $G_{x}$ on each fiber $\Vsheaf(x)$ which
depends differentiably on $x$. The pair $(\Vsheaf,G)$ is often
referred to as an Hermitian vector bundle. 
For a given frame, $e_{a}(x)$, 
the Hermitian structure specifies an inner product as
\begin{equation}
  \label{Gdef} 
  (e_a, e_b)=G_{{\bar a}b}
  ,\qquad
  G=G^{\dagger}
  .
\end{equation}
A choice of frame provides the necessary coordinates to express the
covariant derivative in terms of the connection,
\begin{equation}
  \label{covd} 
 D(v^a e_a) =(dv^a)e_a + v^a A^{b}_{a}e_{b}.
 \end{equation}
 Imposing compatibility of the connection with the holomorphic
 structure of the bundle and the fiber metric determines the
 connection uniquely up to gauge transformations. Written in the most
 useful gauge choice for our purposes, the connection is
\begin{equation}
  \label{Amath} 
  \bar{A}=0
  ,\qquad
  A=G^{-1}\partial G
 .
\end{equation} 
One can then rephrase the Hermitian Yang-Mills equation for
$F^{(1,1)}$ in \eqref{hym_genr} as a condition on the bundle metric,
\begin{equation}
  \label{herm_met} 
  \mu(\Vsheaf)\cdot \mathbf{1}_{n \times n} 
  =
  g^{i\bar{j}}F_{i\bar{j}}
  =
  g^{i\bar{j}}\bar{\partial}_{\bar{j}}A_i
  =
  g^{\bar{j}i}\bar{\partial}_{\bar{j}}(G^{-1}\partial_i G)
 .
\end{equation} 
A metric $G$ on the fiber of $\Vsheaf$ satisfying this equation is
called an ``Hermite-Einstein metric''. By integration, this metric can
be used to define an inner product on the space of global sections of
$\Vsheaf$, $s^{a}_\alpha$ where $\alpha=1,\ldots h^0(X,\Vsheaf)$, 
\begin{equation}
  \label{sec_prod} 
  \big\langle s_{\alpha} \big| s_{\beta} \big\rangle
  =\int_X
  s^{b}_{\beta}G_{b\bar{a}}\bar{s}^{\bar{a}}_{\bar{\alpha}}
  ~\dVol
   .
\end{equation}

The above notions in differential geometry can be related to seemingly very different concepts in the algebraic geometry of holomorphic vector bundles. Relating the two approaches has made it possible to better understand both. For K\"ahler manifolds, the relationship
can be summarized as follows:
\begin{theorem}[Donaldson-Uhlenbeck-Yau~\cite{duy1,duy2}]
  On each slope \emph{poly-stable} holomorphic vector bundle,
  $\Vsheaf$, there exists a unique connection satisfying the general
  Hermitian Yang-Mills equations eq.~\eqref{hym_genr}. Moreover, such
  a connection exists if and only if $\Vsheaf$ is slope poly-stable.
\end{theorem}
Thus, in the heterotic string context, to verify that a gauge vector
bundle is consistent with supersymmetry one need only verify that it
is slope poly-stable. The notion of slope-stability of a bundle $\cF$
over a K\"ahler manifold $X$ is defined by means of a real number (the
same which appeared in eq.~\eqref{hym_genr}), called the \emph{slope}:
\begin{equation}
  \label{slope} 
  \mu (\cF)
  \equiv
  \frac{1}{\rk(\cF)}\int_{X}c_{1}(\cF)\wedge \omega^{d-1} 
  ,
\end{equation} 
where $d$ is the complex dimension of the K\"ahler manifold. Here,
$\omega$ is the K\"ahler form on $X$, while $\rk(\cF)$ and
$c_1(\cF)$ are the rank and the first Chern class of $\cF$
respectively. A bundle $\Vsheaf$ is called \emph{stable}
(\emph{semi-stable}) if, for all sub-sheaves $\cF\subset \Vsheaf$ with
$0<\rk(\cF)<\rk(\Vsheaf)$, the slope satisfies
\begin{equation}
  \label{slope_req} 
  \mu (\cF) 
  ~
  \begin{smallmatrix}
    \displaystyle < 
    \\[1mm]
    (\leq)
  \end{smallmatrix}
  ~
  \mu(\Vsheaf)
  .
\end{equation} 
A bundle is \emph{poly-stable} if it can be decomposed into a direct sum of
stable bundles 
%($\Vsheaf=\bigoplus_n \Vsheaf_n$), 
which all have the same slope. That is, 
%($\mu(\Vsheaf_i)=\mu(\Vsheaf)$). 
%
\begin{equation}
\label{slope_themovie}
\Vsheaf=\bigoplus_n \Vsheaf_n  , \qquad \mu(\Vsheaf_i)=\mu(\Vsheaf)  .
\end{equation}

From the above definitions, it is clear that the condition of
slope-stability on a Calabi-Yau manifold depends on all moduli of the
heterotic compactification. To be specific, consider a Calabi-Yau
threefold. Here, the moduli are the $h^{1,1}(X)$ K\"ahler moduli, the
$h^{2,1}(X)$ complex structure moduli, and the $h^1(\End(\Vsheaf))$
vector bundle moduli. The dependence on K\"ahler moduli is explicit in
eqns.~\eqref{slope} and~\eqref{hym_genr}. Since slope stability is an
open property~\cite{huybrechts}, it depends only on a K\"ahler form,
$\omega$, defined up to an overall scale. We refer to this
one-parameter family of K\"ahler forms (which define a \emph{ray} in
K\"ahler moduli space) as a choice of ``polarization'' and frequently
make no distinction between a particular $\omega$ and its associated
polarization. It is possible to expand the K\"ahler form $\omega$ in
\eqref{slope} as $\omega=t^{r}\omega_{r}$, where $\omega_{r}$ are a
basis of $(1,1)$-forms and $t^r$ are the real parts of the K\"ahler
moduli. Written in terms of the triple intersection numbers $d_{rst}$
of the threefold, the slope is simply
\begin{equation}
  \label{slope2the revenge}
  \mu(\Vsheaf)=\frac{1}{\rk(\Vsheaf)}
  \sum_{r,s,t=1}^{h^{1,1}(X)}
  d_{rst}
  c_1(\Vsheaf)^r t^s t^t   
  .
\end{equation}
The complex structure moduli of the Calabi-Yau manifold and
the vector bundle moduli enter through the notion of a subsheaf $\cF \subset \Vsheaf$. Thus,
finding a solution to the Hermitian Yang-Mills equations, or
determining whether the bundle is slope-stable, is a question that
must be asked after selecting a particular point in moduli space.

%%%%%%%%%%%%%%%%%%%%%%%%%%%%%%%%

\section{The Generalized Donaldson Algorithm}
\label{metric_review}

Many of the challenges associated with string compactifications on a
Calabi-Yau $d$-fold $X$ arise from the difficulty in determining the
explicit geometry. The simplest $\Ncal=1$ supersymmetric vacuum
solutions require a Ricci-flat metric, $g_{i\bar{j}}$, on $X$ and a
Hermite-Einstein bundle metric, $G_{\bar{a}b}$, satisfying \eqref{herm_met} as
discussed above. While Yau's theorem~\cite{MR480350} ensures that a
Ricci flat metric exists on a Calabi-Yau manifold, and the
Donaldson-Uhlenbeck-Yau theorem~\cite{duy1,duy2} provides for the
existence of a Hermite-Einstein metric on a slope-stable bundle, no
analytic solutions for either the metric or connection have yet been
found.

However, recent work has made it possible to find accurate numerical
solutions for both metrics and connections. An algorithm was initially
proposed by Donaldson for the computation of Ricci-flat
metrics~\cite{MR2161248, MR1916953, DonaldsonNumerical}, and was
implemented numerically and extended in~\cite{MR2283416, Braun:2007sn,
  Braun:2008jp, Douglas:2006rr, Douglas:2006hz,
  Headrick:2009jz,Anderson:2010ke}. What we refer to as the
``generalized Donaldson algorithm'' is an extension of Donaldson's
approximation scheme which numerically approaches an Hermite-Einstein
bundle metric, solving \eqref{herm_met}. This was developed
mathematically in~\cite{MR1064867,MR2154820} and implemented
numerically in~\cite{Douglas:2006hz,Anderson:2010ke}. A thorough
review of the Donaldson algorithm and its extensions is beyond the
scope of this paper. We refer the reader to~\cite{Anderson:2010ke} for
more details. However, in order to proceed with our present
investigation of K\"ahler cone substructure, we provide here a brief
review of the central ingredients of the (generalized) Donaldson's
algorithm and set the notation that will be used throughout this work.

\subsection{Donaldson's Algorithm}\label{metric_alg}

We begin with an overview of Donaldson's algorithm for approximating
the Ricci flat metric on a Calabi-Yau manifold. The first ingredient
we need is one of the simplest K\"ahler metrics, the Fubini-Study
metric on $\mathbb{P}^n$. This is given by $g_{FS
  i\bar{j}}=\frac{i}{2}\partial_{i} \bar{\partial}_{\bar{j}}K_{FS}$,
where
\begin{equation}
  \label{FS} 
  K_{FS} = \frac{1}{\pi}  \ln \sum_{i\bar{j}}
  h^{i\bar{j}}z_{i}\bar{z}_{\bar{j}}
\end{equation} 
and $h^{i\bar{j}}$ is any Hermitian, positive, non-singular matrix. 

Since it is always possible to embed $X\subset \mathbb{P}^n$ for some
large enough $n$, the Fubini-Study metric can be used to induce some
metric on any Calabi-Yau manifold $X$. Such a metric will not be
Ricci-flat, for otherwise one could easily write down an analytic
expression for the Calabi-Yau metric. It is tempting to wonder whether
there exists a generalized version of eq.~\eqref{FS} with enough free
parameters to provide a more versatile induced metric on $X$?  The
central idea of Donaldson's algorithm is to find such a generalization
and a procedure for successively tuning its free parameters to
approximate the Ricci-flat metric. The obvious generalization of
eq.~\eqref{FS} is to replace the degree one polynomials with polynomials
of higher degree. That is,
\begin{equation}
  \label{FS_gen} 
  K = \frac{1}{k\pi} ~\ln\!\! \sum_{i_{1}\ldots
    i_{k}\bar{j}_{1}\ldots\bar{j}_{k}} h^{i_{1}\ldots
    i_{k}\bar{j}_{1}\ldots\bar{j}_{k}}z_{i_{1}}\ldots
  z_{i_{k}}\bar{z}_{\bar{j}_{1}}\ldots \bar{z}_{\bar{j}_{k}}
\end{equation} 
where $h^{i_{1}\ldots i_{k}\bar{j}_{1}\ldots\bar{j}_{k}}$ is
Hermitian. This new K\"ahler potential now has $(n+1)^{2k}$ real
parameters. This generalization can, in fact, be seen in a more
systematic way by using holomorphic line bundles over $X$. The Kodaira
Embedding Theorem~\cite{MR507725} tells us that given an ample
holomorphic line bundle $\Lsheaf$ over $X$ with $n_{1}=h^0(X,
\Lsheaf)$ global sections, one can define an embedding of $X$ into
projective space via the sections of $\mathcal{L}^{k}
=\Lsheaf^{\otimes k}$ for some $k$. That is, choosing a basis for the
space of sections, $s_{\alpha} \in H^0(X,\Lsheaf^k)$ where $0 \leq
\alpha \leq n_{k}-1$, allows one to define a map from $X$ to
$\mathbb{P}^{n_{k}-1}$ given by
\begin{equation}
  \label{embed} 
  i_{k}:~ X \to \mathbb{P}^{n_{k}-1} 
  ,\quad
  (x_{0},\ldots,x_{d-1}) \mapsto \big[ s_0(x): \ldots : s_{n_{k}-1}(x) \big]
   ,
\end{equation} 
where $x_i$ are holomorphic coordinates on the Calabi-Yau manifold. If
$\mathcal{L}$ is sufficiently ample, eq.~\eqref{embed} will define an
embedding of $ X \subset \mathbb{P}^{n_{k}-1}$ for all $\mathcal{L}^k$
with $k \geq k_0$ for some $k_0$.

In terms of this embedding via a line bundle $\cL$, one can view the
generalized K\"ahler potential in eq.~\eqref{FS_gen}, restricted to
$X$, as simply
\begin{equation}
  \label{Lmetric} 
  K_{h,k}
  =\frac{1}{k\pi} \ln \sum_{\alpha,\bar{\beta}=0}^{n_{k}-1} h^{\alpha\bar{\beta}}
  s_{\alpha}{\bar{s}}_{\bar{\beta}}=\ln ||s||^{2}_{h,k}  .
\end{equation} 
Geometrically, \eqref{Lmetric} defines an Hermitian
fiber metric on the line bundle $\Lsheaf^{k}$ itself. It provides a natural
inner product on the space of global sections
\begin{equation}
  \label{sec_metric}
  M_{\alpha\bar{\beta}}=\left<s_{\beta}|s_{\alpha}\right> 
  = 
  \frac{n_{k}}{\Vol_{CY}(X)} \int_{X}
  \frac{s_{\alpha}{\bar{s}}_{\bar{\beta}}}{||s||^{2}_{h}}\dVol_{CY}  ,
\end{equation} 
where
\begin{equation}
  \label{metric_themovie}
  \dVol_{CY}=\Omega \wedge {\bar{\Omega}}
\end{equation}
and $\Omega$ is the holomorphic (3,0) volume form on $X$. 

With the initial K\"ahler metric eq.~\eqref{Lmetric} in hand, we must
now proceed to systematically adjust it towards Ricci flatness. To
accomplish this, the notion of a balanced metric is required. Note
that, in general, the matrices $h^{\alpha\bar{\beta}}$ and
$M_{\alpha\bar{\beta}}$ in \eqref{sec_metric} are completely
unrelated. However, for special metrics, they may coincide.  The
metric $h$ on the line bundle $\Lsheaf$ is called \emph{balanced} if
\begin{equation} 
  (M_{\alpha\bar{\beta}})^{-1}
  =
  h^{\alpha\bar{\beta}} 
  .
\end{equation} 
Donaldson first recognized that balanced metrics lead to special
curvature properties. These can be summarized as
follows~\cite{MR2161248, MR1916953, DonaldsonNumerical,
  2007arXiv0709.1490K}:
\begin{theorem}[Donaldson, Keller]
  For each $k \geq 1$, the balanced metric $h$ on $\Lsheaf^k$ exists
  and is unique. As $k \to \infty$, the sequence of metrics
  \begin{equation}
    \label{g_approx}
    g_{i\bar{j}}^{(k)}
    =
    \frac{1}{k\pi}\partial_{i} \bar{\partial}_{\bar{j}}
    \ln \sum_{\alpha,\bar{\beta}=0}^{n_{k}-1}
    h^{\alpha\bar{\beta}}s_{\alpha}\bar{s}_{\bar{\beta}}
  \end{equation} 
  on $X$ converges to the unique Ricci-flat metric for the given
  K\"ahler class and complex structure.
\end{theorem}

The central task of Donaldson's algorithm is thus to find the balanced
metric for each $k$. To this end, Donaldson defined the T-operator as
\begin{equation}
  \label{t_oper}
  T(h)_{\alpha\bar{\beta}}
  =
  \frac{n_{k}}{\Vol_{CY}(X)}
  \int_{X}
  \frac{s_{\alpha}{\bar{s}}_{\bar{\beta}}}{\sum_{\gamma\bar{\delta}}
    h^{\gamma\bar{\delta}}s_{\gamma}{\bar{s}}_{\bar{\delta}}}\dVol_{CY}
  .
\end{equation} 
For a given metric $h$, it computes a matrix $T(h)$. If this matrix
equals $M_{\alpha\bar{\beta}}$, we have a balanced embedding. To find
this fixed point, simply iterate \eqref{t_oper} as follows.
\begin{theorem}
  For any initial metric $h_0$ (and basis $s_{\alpha}$ of
  global sections of $\Lsheaf^k$), the sequence
  \begin{equation} 
    h_{m+1}=\big(T(h_m)\big)^{-1}
  \end{equation} 
  converges to the balanced metric as $m \to \infty$.
\end{theorem} 
\noindent Happily, in practice, very few ($\approx 10$) iterations are
needed to approach the fixed point. Henceforth, we will also refer to
$g^{(k)}_{i\bar{j}}$ in eq.~\eqref{g_approx}, the approximating metric
for fixed $k$, as the balanced metric. It should be noted that, to
find the balanced metric at each step $k$, one must be able to
integrate over the Calabi-Yau threefold. In \autoref{volker_int}, we
will discuss the new adaptive mesh numerical integration scheme used
throughout this work.

As one final ingredient in the algorithm, one must be able to quantify
how closely the numerical metric approximates the Ricci-flat metric. A
variety of such error measures were given
in~\cite{Anderson:2010ke}. Recall that, given an sufficiently ample
line bundle $\mathcal{L}$, one can find a K\"ahler form
\begin{equation}
  \label{kform}
  \omega_k
  =
  \frac{i}{2}g^{(k)}_{i\bar{j}}dz_{i}\wedge d\bar{z}_{\bar j}
\end{equation} 
corresponding to the balanced metric associated with
$\mathcal{L}^k$. Note that the K\"ahler class of this K\"ahler form is
$[\omega_k]=2\pi c_{1}(\mathcal{L}^k)$ and the associated volume is
\begin{equation}\label{voldef}
  \Vol_k = \frac{1}{d!}\int_X \omega_k^d
  ,
\end{equation} 
where $\omega_{k}^{d}$ denotes the $(d,d)$ volume form $\omega \wedge \dots \wedge \omega$.

In this paper, we will measure convergence of the Donaldson algorithm via the Ricci scalar in
\begin{equation}
   ||EH||_k = 
   \Vol_k^{(1-d)/d} 
    \int |R_{k}| \; \sqrt{\det g_{k}} \; \text{d}^{2d}\! x 
    .
\label{again2}
\end{equation}
On a Calabi-Yau manifold, $||EH||_k =O(k^{-1})$ as $k\to \infty$ and,
hence, this error measure should approach zero. As a final note, we
will henceforth denote the degree of twisting, given by the
integer $k$ in $\cL^{k}$, as $k_g$ to make it clear that this integer
is associated with the computation of the metric.

A summary of Donaldson's algorithm for Ricci flat metrics is provided in \autoref{algorithm_review}. We now turn to the generalized Donaldson algorithm for computing Hermite-Einstein fiber metrics on holomorphic vector bundles.

%%%%%%%%%%%%%%%%%%%%%%%%
\subsection{Hermite-Einstein Bundle Metrics}\label{connection_alg}

As we saw in the previous section, Donaldson's algorithm is a powerful
tool for numerically approximating the Ricci-flat metric on a
Calabi-Yau manifold. In this section, we investigate a generalization
of these techniques which can be used to approximate the field
strength $F^{(1,1)}$ of a holomorphic connection which satisfies
\eqref{hym_genr}. As discussed in \autoref{metric_alg}, Donaldson's
algorithm for Calabi-Yau metrics can be viewed as a method for
numerically obtaining a particular Hermitian structure on the ample
line bundle $\cL^k$. This balanced fiber metric on $\cL^k$ allows one
to define a balanced embedding of the Calabi-Yau space $X$ into
$\mathbb{P}^{n_{k}-1}$.  By mapping the coordinates $x \in X$ into the
global sections $s_{\alpha} \in H^0(X,\cL^{k})$, that is,
\begin{equation}
  \xymatrix@R=5mm@C=2cm{
    x \ar@{|->}[r] & 
    [ s_0(x) : \cdots : s_{n_k-1}(x)]  ,
  }
\end{equation}
we produced a map $i_{k}: X \to \mathbb{P}^{n_{k}-1}$ where
$n_k=h^0(X,\cL^k)$. The pull-back of the associated Fubini-Study
metric was shown in \autoref {metric_alg} to converge to the
Ricci-flat metric on $X$ in the limit that $k \to \infty$. Viewed in
terms of Hermitian fiber metrics on line bundles, it is a natural
question to ask whether Donaldson's algorithm could be extended to
develop an analogous approximation to Hermitian metrics on higher rank
vector bundles. In particular, could one find an approximation scheme
to produce an Hermitian metric on an arbitrary stable bundle $\Vsheaf$
of rank $n$ such that it satisfies condition \eqref{herm_met}?
Fortunately, precisely this question has been addressed in the
mathematics literature~\cite{MR2154820} and implemented for physics
in~\cite{Douglas:2006hz, Anderson:2010ke}.

To generalize Donaldson's algorithm, consider defining an embedding
via the global sections of a twist of some holomorphic vector bundle
$\Vsheaf$ with non-Abelian structure group. That is, consider a map
\begin{equation}\label{bundle_embed}
  \xymatrix@R=5mm@C=2cm{
    x \ar@{|->}[r] & 
    {
      \left[
        \begin{pmatrix}
          S_0^1(x) \\ \vdots \\ S_0^n(x)     
        \end{pmatrix}
        : \cdots :
        \begin{pmatrix}
          S_{N_k-1}^1(x) \\ \vdots \\ S_{N_k-1}^n(x)
        \end{pmatrix}
      \right]
      .
    }}
\end{equation}
from $x \in X$ into the global sections $S^{a}_{\alpha} \in
H^0(X,\Vsheaf\otimes\cL^k)$, where $\alpha=0 \ldots N_k-1$ indexes the
$h^0(X,\Vsheaf\otimes \cL^k)$ global sections and the index $a=1,\ldots
n$ is valued in the fundamental representation of structure group $K
\subseteq U(n)$ of the rank $n$ bundle $\Vsheaf$. We hope then to
define the embedding
\begin{equation}
  \label{V_embed}
  X \longrightarrow G(n,N_{k}-1)  ,
\end{equation}
where $G(n,N_{k}-1)$ denotes the Grassmannian of the relevant
dimension\footnote{In this language, the Abelian case in \eqref{embed}
  is simply an embedding $X \rightarrow G(1,n_{k}-1)$.}. By the
Kodaira embedding theorem~\cite{MR507725}, given a holomorphic vector
bundle, $\Vsheaf$, and an ample line bundle, $\cL$, there must exist a
finite integer $k_0$ such that, for any $k>k_0$, the twisted bundle
$\Vsheaf (k)=\Vsheaf\otimes \cL^k$ defines an embedding, $i_k: X \to
G(n,N_k-1)$.

As in the Abelian case in the previous section, one can attempt to use
this embedding to define a Hermite-Einstein bundle metric on $\Vsheaf
\otimes \cL^k$ and, hence, an Hermitian Yang-Mills connection as in
\eqref{Amath} and \eqref{herm_met}. If $\cL$ is ample then, for some
sufficiently large $k$, $\Vsheaf \otimes \cL^k$ will be generated by
its global sections. That is, it will define an embedding as in
\eqref{V_embed}. In our search for a solution to the Hermitian
Yang-Mills equation \eqref{hym_genr}, the connection on the twisted
bundle $\Vsheaf \otimes \cL^{k}$ will be closely related to the
original connection, since such a twist only modifies the trace part
of the field strength. Stated in terms of algebraic geometry, the
process of twisting will not modify the slope-stability properties of
$\Vsheaf$ since $\Vsheaf\otimes \cL^k$ is stable if and only if
$\Vsheaf$ is.

As at the beginning of \autoref{metric_alg}, where we chose the trial
form of the K\"ahler potential in \eqref{Lmetric}, here we begin with
another simple anzatz for the Hermitian structure $G$ in
eq.~\eqref{Gdef}. Consider the matrix
\begin{equation}
  \label{G_anzatz}
  (G^{-1})^{a\bar{b}}
  =
  \sum_{\alpha,\beta=0}^{N_k-1}H^{\alpha\bar{\beta}}S_{\alpha}^{a}
  (\bar{S})_{\bar{\beta}}^{\bar{b}}   
  ,
\end{equation}
where $H^{\alpha\bar{\beta}}$ is a Hermitian matrix of constants and
$S^{a}_{\alpha}$ are the global sections of $\Vsheaf\otimes \cL^k$. As
in \eqref{sec_prod}, this fiber metric induces an inner product on the
space of sections $H^0(X, \Vsheaf\otimes \cL^k) =
\Span\{S_{\alpha}\}$ via
\begin{equation}
  \big< S_{\beta} \big| S_{\alpha} \big>=\frac{N_k}{\Vol_{CY}}
  \int_{X}
  S_{\alpha}^{a}(G^{a\bar{b}})^{-1}\bar{S}^{\bar{b}}_{\bar{\beta}}\dVol_{CY}
  =\frac{N_k}{\Vol_{CY}}\int_{X} 
  S_{\alpha}^{a}
  (S^{a}_{\gamma}H^{\gamma\bar{\delta}}{\bar
    S}^{\bar{b}}_{\bar{\delta}})^{-1}
  \bar{S}^{\bar{b}}_{\bar{\beta}}\dVol_{CY}
  .
\label{coffee2}
\end{equation}
With this definition of the inner product on sections, one can give a
natural generalization of the T-operator eq.~\eqref{t_oper}. This
generalization,
\begin{equation}\label{t_gen}
  T(H)_{\alpha\bar\beta} =\frac{N_k}{\Vol_{CY}}
  \int_X       
  S_\alpha 
  \Big( S^\dagger H S \Big)^{-1}
  \bar S_{\bar \beta}
  ~\dVol_{CY}  ,
\end{equation}
was introduced in~\cite{MR2154820} and studied numerically
in~\cite{Douglas:2006hz,Anderson:2010ke}.  Note that if $\Vsheaf$ is a
line bundle then eq.~\eqref{t_gen} reduces to \eqref{t_oper} and one
recovers the case of a balanced embedding into
$\mathbb{P}^{N_{k}-1}$. As in the previous section, we will now
describe how the iteration of the generalized T-operator can produce a
fixed point which describes an Hermite-Einstein bundle metric.

To do this however, we must introduce one additional notion of
stability, namely that of ``Gieseker
stability''~\cite{huybrechts}. Let $\cL$ be an ample line bundle and
$\cF$ be a torsion-free sheaf. The Hilbert polynomial of $\cF$ with
respect to $\cL$ is defined as
\begin{equation}\label{gieseker}
  p_{\cL}(\cF)(n)=\frac{\chi(\cF\otimes \cL^{n})}{\rk(\cF)}
\end{equation} 
where $\chi(\cF\otimes \cL^{n})$ is the index of $\cF\otimes
\cL^{n}$. Given two polynomials $f$ and $g$, we will write $f \prec g$
if $f(n) < g(n)$ for all $n \gg 0$. Then a bundle $\Vsheaf$ is said to
be \emph{Gieseker stable} if, for every non-zero torsion free subsheaf
$\cF \subset \Vsheaf$,
\begin{equation}\label{gieseker_ineq}
  p_{\cL}(\cF) \prec
  p_{\cL}(\Vsheaf)
  .
\end{equation}
With this definition in hand, it was shown in~\cite{MR2154820} that 
\begin{theorem}[Wang]\label{wang2}
  A bundle $\Vsheaf$ is Gieseker stable if and only if the $k$-th
  embedding, defined by $\Vsheaf\otimes \cL^{k}$ as in
  eq.~\eqref{V_embed}, can be moved to a ``balanced'' place. That is,
  if there exists an orthonormal section-wise metric on the twisted
  bundle such that
  \begin{equation}
    \label{T_gen_bal}
    \big( T(H)_{\alpha\bar\beta} \big)^{-1} 
    =
    H^{\alpha\bar{\beta}}
  \end{equation}
  is a fixed point of the generalized T-operator .
\end{theorem}

We can use this special metric on $\Vsheaf \otimes \cL^k$ to define an
Hermitian metric on $\Vsheaf$ itself. Let $G_{\cL}$ denote the
balanced metric on $\cL$, and $G^{(k)}$ the balanced metric on $\Vsheaf
\otimes \cL^{k}$. Then
\begin{equation}
  \label{tensor_met}
  G_{k}=G^{(k)}\otimes G_{\cL}^{-k}
\end{equation}
is an Hermitian metric on $\Vsheaf$. This appears in the following
important theorem~\cite{2008arXiv0804.4005S, MR2154820, MR2180559}.
\begin{theorem}[Seyyedali, Wang]
  \label{wang1}
  Suppose $\Vsheaf$ is a Gieseker stable bundle of rank $n$. If $G_k
  \to G_{\infty}$ as $k \to \infty$, then the metric $G_\infty$ solves
  the ``weak Hermite-Einstein equation''
  \begin{equation}
    \label{weak_hym}
    g^{i \bar j} F_{i \bar j}
    =
    \left( 
      \mu + \frac{\overline{R}-R}{2}
    \right)
    \mathbf{1}_{n\times n}
  \end{equation}
  where
  \begin{itemize}
  \item $R$ is the scalar curvature.
  \item $\overline{R} = \int R \sqrt{\det g} \; \text{d}^{2d}\!x$ is
    the averaged scalar curvature, which is zero for any K\"ahler
    metric on a manifold of vanishing first Chern class.
  \end{itemize}
\end{theorem}

We will, henceforth, denote the degree $k$ of the embedding defined
above as $k_{H}$, to make clear its association with the Hermitian
matrix in eq.~\eqref{G_anzatz} and distinguish it from
$k_g$. Procedurally, the process of obtaining the Hermite-Einstein
fiber metric on a slope-stable bundle $\Vsheaf$ is very similar to
that outlined for the Ricci-flat connection in \autoref{metric_alg}:
for each value $k_{H}$ of the twisting, we iterate the T-operator
associated with the embedding defined by $H^0(X,\Vsheaf \otimes
\cL^{k_{H}})$ until a fixed point is reached. Then, by
\autoref{wang1}, the induced connection approximates solutions to
eq.~\eqref{weak_hym} as $k_{H} \to \infty$. However, there is an
immediate and important difference between this generalized algorithm
and Donaldson's algorithm for Ricci-flat metrics. While all Calabi-Yau
manifolds admit a Ricci-flat metric, not all holomorphic vector
bundles will admit an Hermite-Einstein metric satisfying
\eqref{herm_met}. That is, if one applies the algorithm to a bundle
that is not slope-stable, it will not converge to a solution of the
Hermitian Yang-Mills equations, \eqref{hym_genr}. Moreover, it should
be noted that while all slope-stable bundles are Gieseker
stable~\cite{huybrechts}, the converse does not hold: not all
Gieseker-stable bundles are slope-stable. That is, there exist cases
where the iteration of the T-operator does converge for fixed $k_{H}$,
but the sequence of metrics does not converge towards a solution of
the Hermite-Einstein fiber metric. However, if $\Vsheaf$ is a
slope-stable holomorphic bundle, then the iteration $H_{m+1} =
T(H_m)^{-1}$ \emph{will} converge at each $k_{H}$, and in the limit
that $k_{H} \to \infty$, produce the Hermitian bundle metric
$G_{\infty}$ satisfying \eqref{weak_hym} via its associated field
strength defined in \eqref{Amath} and \eqref{herm_met}. Moreover, in
the case where the Calabi-Yau metric $g^{i \bar j}$ is Ricci-flat,
\eqref{weak_hym} simply reduces to \eqref{hym_genr}. Thus, we have
found a solution to the Hermitian Yang-Mills equations.

However, one must be careful. Despite having found a Hermite-Einstein
bundle metric (and, hence, HYM connection) associated with the twisted
bundle $\Vsheaf\otimes \cL^{k_H}$, and an Hermitian metric
$G_{\infty}$ satisfying \eqref{weak_hym}, our task is not yet
complete. We still need to explicitly determine the connection on the
bundle $\Vsheaf$ itself satisfying the Hermitian Yang-Mills equations,
\eqref{hym_genr}. Since the process of twisting $\Vsheaf$ by a line
bundle $\cL^{k_{H}}$ in the above construction clearly modifies the
trace-part of the connection, one must subtract this line bundle
contribution to get the connection on $\Vsheaf$ \emph{only}. To do
this, we have to separately find a suitable metric $G_{\cL}$ on
$\Lsheaf$. For example, one could compute the balanced metric
$G_\cL^{(k_h)} = s^\dagger h s$ on $\Lsheaf^{k_h}$ for some
sufficiently large $k_h$. Then $G_{\cL}=(s^\dagger h s)^{1/k_{h}}$
would approximate the constant curvature Hermitian fiber metric on
$\Lsheaf$ and, as in eq.~\eqref{tensor_met}, we find that
\begin{equation}
  G
  =
  G^{(k_H)}\times G_{\cL}^{-k_H}
  =
  \Big( S^\dagger H S\Big) \Big( s^\dagger h s\Big)^{-k_H/k_h}
\end{equation}
is the fiber metric \eqref{Gdef} on $\Vsheaf$. As before, $S\in
H^0(X,\Vsheaf \otimes \cL^{k_H})$ and $s \in H^0(X, \cL^{k_h})$ are
the relevant global sections. Using eqns.~\eqref{Amath} and
\eqref{herm_met}, in terms of the Hermitian metric, the connection on
$\Vsheaf$ is then given by
\begin{equation}\label{Auntwist}
  \begin{split}
    A(\Vsheaf) 
    =&\;
    \partial
    \left[ 
      \Big( S^\dagger H S\Big) 
      \Big( s^\dagger h s\Big)^{-k_H/k_h}
    \right]
    \Big( S^\dagger H S\Big)^{-1}
    \Big( s^\dagger h s\Big)^{k_H/k_h}
    \\=&\;
    A(\Vsheaf \otimes \Lsheaf^{k_H})
    - \frac{k_H}{k_h}
    A(\Lsheaf^{k_h})  .
  \end{split}
\end{equation}
That is, one can ``untwist'' the connection simply by subtracting the
trace of the Abelian connection on
%$\Vsheaf \otimes 
$\cL^{k_{H}}$ to produce the $U(n)$ connection on $\Vsheaf$. The curvature is given by
\begin{equation}
  \label{eq:Funtwisted}
  F^{(0,2)} = F^{(2,0)} = 0
  ,\quad
  g^{i \bar j} F_{i \bar j}  = g^{i \bar j} \partial_{\bar j} A_i
  =
  g^{i \bar j} \partial_{\bar j} \partial_i \ln 
  \Big( S^\dagger H S\Big) \Big( s^\dagger h s\Big)^{-k_H/k_h} .
\end{equation}    

As shown in~\cite{Anderson:2010ke}, when $\Vsheaf$ is a $U(n)$ bundle,
the most efficient way to perform this untwisting is not by computing
an independent balanced metric $(G_{\cL})^{k_h}$, but directly using
the induced Hermitian fiber metric on the determinant line bundle
$\wedge^{n}(\Vsheaf\otimes\Lsheaf^{k_H})$ of $\Vsheaf\otimes
\cL^{k_{H}}$. In particular, we choose $k_h = \rank(\Vsheaf) k_H$. It
follows that the Hermitian metric on $\cL^{k_{h}}$ is
\begin{equation}
  ( G_{\cL})^{k_h} 
  =
  \det(G^{(k_H)})
  =
  \det \big( S^\dagger H  S\big)  .
\end{equation}

Let $\lambda^{(k_H)}$ be the eigenvalues of $g^{i\bar
  j}F^{(k_H)}_{i\bar j}$ on $\Vsheaf\otimes \Lsheaf^{k_{H}}$, and let
$\lambda$ be the corresponding eigenvalues of $ g^{i\bar j}F_{i\bar
  j}$ on $\Vsheaf$ after untwisting. Using eq.~\eqref{eq:Funtwisted},
we obtain that
\begin{equation}\label{lambdai}
  \lambda_i 
  = \lambda^{(k_H)}_i - 
  \frac{1}{\rank \Vsheaf} g^{i\bar j} \tr F^{(k_H)}_{i\bar j}
  = \lambda^{(k_H)}_i - \frac{\sum_j
    \lambda^{(k_H)}_j}{\rank \Vsheaf}  ,
\end{equation}
where $i=1,\ldots n$ where $n$ is the dimension of the fundamental
representation of the structure group of $\Vsheaf$. Therefore, the
effect of this untwisting is precisely to subtract, at each point, the
average of the eigenvalues. Hence, in~\cite{Anderson:2010ke} we
referred to this untwisting as subtracting the trace.

The eigenvalues in \eqref{lambdai} are a pointwise measure of the
error in the numerically derived connection. For a slope-stable
bundle, $\lambda_i \to \mu$ as one increases $k_{H} \to \infty$. To
properly define an error measure for the approximation to the
Hermitian Yang-Mills connection, we must test the approximation at all
points and, hence, integrate \eqref{lambdai} over $X$. As
in~\cite{Anderson:2010ke}, we define the $L^1$ error measure
\begin{equation}
  \label{eq:taudef}
  \tau(A_\Vsheaf) = 
  \frac{1}{2\pi}
  \frac{
    k_g
  }{
    \Vol_{k_{g}}
    \rank(\Vsheaf)
  }
  \int_X \Big( \sum |\lambda_i| \Big) 
  \sqrt{g} \mathop{d}\nolimits^{2d}\!x  ,
\end{equation}
where $\Vol_{k_{g}}$ is the volume computed in \eqref{voldef}. For a
slope-stable bundle, $ \tau(A_\Vsheaf) \to \rank(\Vsheaf) \mu$ as
$k_{H} \to \infty$. This is simply a global check of the eigenvalues
in \eqref{lambdai}. To summarize the results of this section, the
generalized Donaldson algorithm for numerically approximating a
Hermitian Yang-Mills connection is presented in
\autoref{algorithm_review}.
%%%%%%%%%%%%%%%%%%%%%%%%%%%%%%%%%%%%%%%%%%%%%%%%
\begin{sidewaystable}[htbp]
  \begin{center}
    \renewcommand{\arraystretch}{2}
    \begin{tabular}{|c|p{0.4\linewidth}|p{0.4\linewidth}|}
      \hline
      \textbf{Step} & 
      \textbf{Ricci-flat metric on \boldmath$X$} & 
      \textbf{Hermite-Einstein metric on \boldmath$\Vsheaf$}
      \\  \hline\hline
      1 
      & 
      Choose an ample line bundle $\cL$ and a degree $k_g$. &
      Choose an ample line bundle $\cL$, a degree $k_{H}$ and form the
      twisted bundle $\Vsheaf \otimes \cL^{k_H}$.
      \\  \hline
      2 
      & 
      Find a basis $\{s_{\alpha}\}_{\alpha=0}^{n_{k}-1}$ for $H^0(X,\mathcal{L}^{k_g})$
      at the chosen $k_g$.
      &
      Find a basis $\{S_{\alpha}\}_{\alpha=0}^{N_{k_{H}}-1}$ for $H^0(X,\Vsheaf \otimes \cL^{k_{H}})$
      at the chosen $k_{H}$.
      \\  \hline
      3 &
      Choose an initial positive, Hermitian matrix
      $h^{\gamma\bar{\delta}}$ for the ansatz eq.~\eqref{Lmetric}. Numerically
      integrate to compute the T-operator in eq.~\eqref{t_oper}.
      &
      Choose an initial positive, Hermitian matrix,
      $H^{\gamma\bar{\delta}}$ for the ansatz \eqref{G_anzatz}. Numerically
      integrate to compute the T-operator in \eqref{t_gen}.
      \\ \hline
      4 
      &
      Set the new $h^{\alpha\bar{\beta}}$ to be
      $h^{\alpha\bar{\beta}}=(T_{\alpha\bar{\beta}})^{-1}$.
      &
      Set the new $H^{\alpha\bar{\beta}}$ to be
      $H^{\alpha\bar{\beta}}=(T_{\alpha\bar{\beta}})^{-1}$.
      \\ \hline
      5 &
      Return to item 3 and repeat until
      $h^{\alpha\bar{\beta}}$ approaches its fixed point ($\approx$ $10$ iterations).
      &
      Return to item 3 and repeat until
      $H^{\alpha\bar{\beta}}$ approaches its fixed point ($\approx$ $10$ iterations).
      \\ \hline
      6 & &
      Compute the ``untwisted'' connection and
      field strength via \eqref{Auntwist} and \eqref{eq:Funtwisted}.
      \\ \hline
      7 & 
      Measure the error $||EH||_{k_{g}}$.
      & 
      Measure the error $\tau(A_\Vsheaf)_{k_{H}}$.
      \\ \hline
  \end{tabular}
  \caption{The Donaldson algorithm and the generalized Donaldson
    algorithm. In the first column is an outline of the original
    algorithm for the computation of the Ricci-flat metric on a
    Calabi-Yau manifold, $X$. In the second column is an outline of
    the generalized Donaldson algorithm for numerically approximating
    a Hermite-Einstein bundle metric on a slope-stable bundle,
    $\Vsheaf$, over $X$.}
  \label{algorithm_review}
\end{center}
\end{sidewaystable}

%%%%%%%%%%%%%%%%%%%%%%%%%%%%%%%%%%
\section{K\"ahler Cone Substructure}\label{substruc_sec}

\subsection{Modifications For Higher Dimensional K\"ahler Cones}
\label{modific}

One of our central motivations in this work is to understand the
generalized Donaldson algorithm on manifolds with higher dimensional
K\"ahler cones, $\Kcone$, that is, $\dim (\Pic(X))>1$. In particular, we
will compare the behavior of bundles under the algorithm for
\emph{different choices of polarization}. In general, holomorphic
vector bundles can display different slope-stability properties for
different choices of polarization, that is, along different rays in
the K\"ahler cone.\footnote{The radial direction along a fixed ray
  only parametrizes the overall volume and does not change the
  stability properties.} That is, a given bundle may be slope-stable
in some sub-cone $\Kcone_\text{stable} \subset \Kcone$, but be
slope-unstable (and, hence, break supersymmetry) in other sub-regions
$\Kcone_\text{unstable} \subset \Kcone$. This substructure is of
interest both mathematically and physically, with applications ranging
from supersymmetry breaking in heterotic $\Ncal=1$ supersymmetric
vacua \cite{Sharpe:1998zu, Anderson:2009sw, Anderson:2009nt,
  Anderson:2010tc, Anderson:2010mh, Anderson:2011cz} to the
computation of Donaldson-Thomas invariants on Calabi-Yau threefolds
\cite{Thomas:math9806111, 2010arXiv1002.4080L, Anderson:2010ty}. In
general, it is a difficult task to determine the global
slope-stability properties of a vector bundle, $\Vsheaf$, throughout the
K\"ahler cone. In particular, this analysis scales badly with the
dimension of K\"ahler moduli space, $h^{1,1}(X)$. Already for
$h^{1,1}>4$ it becomes prohibitively difficult to analytically analyze
the stability of a bundle except in special cases. As a result, it is
of considerable interest to ask the question: \emph{Can the
  generalized Donaldson algorithm provide an efficient probe of
  K\"ahler cone substructure and vector bundle stability for higher
  dimensional K\"ahler cones?} In principle, the connection algorithm
reviewed in \autoref{connection_alg} shows no difference in
computational difficulty for \emph{any dimension $h^{1,1}$}. That is,
it depends only on a one-dimensional ray (defined by the line bundle
$\cL$ in Step $1$ of \autoref{algorithm_review} and the embedding
\eqref{bundle_embed}) and not on the dimension of the K\"ahler cone
containing that ray. As we will see in the following sections, the
generalized Donaldson algorithm does indeed provide a powerful new
tool for analyzing K\"ahler cone substructure.

In order to pursue this goal, however, one will need a way to compare
the convergence of the algorithms (for both metric and connection) for
different rays in K\"ahler moduli space. A number of properties change
in the case that $h^{1,1}>1$ and, in particular, a few of the
definitions introduced in \autoref{review_section}, and in the
previous literature~\cite{Douglas:2006hz,Anderson:2010ke}, need some
modifications in order to make sensible comparisons for different
polarizations.  One of the first of these is the way in which we
measure the complexity of the embeddings in \eqref{embed} and
\eqref{bundle_embed}. Recall that the algorithms described in
\autoref{review_section} rely on defining an embedding into some
high-dimensional Grassmanian, \eqref{bundle_embed}. For example, to
compute the Ricci-flat metric of $X$, we define the embedding $X \to
\mathbb{P}^n$ via the global sections $H^0(X,\cL^{k_{g}})$. For a
manifold with $h^{1,1}=1$, it is clear that as we increase the degree,
$k_g$, of twisting, we increase the number of global sections and,
hence, as described in \autoref{metric_alg}, the accuracy of the
metric approximation. For example, in~\cite{Anderson:2010ke} we
computed Ricci-flat metrics on the Quintic hypersurface in
$\mathbb{P}^4$, where the global sections of the embedding line
bundle, $\cL=\cO(1)$, increase with $k_g$ as $H^0\big(X,
\cO(k_g)\big)=\frac{5}{6}(5k_g +{k_{g}^3)}$.

However, for manifolds with $h^{1,1}>1$ the situation becomes more
subtle if one wants to compare results for two different
polarizations, defined by line bundles $\cL_1$ and $\cL_2$. As an
example, consider the Calabi-Yau $3$-fold $X$ defined as a $(2,4)$
hypersurface in $\mathbb{P}^1 \times \mathbb{P}^3$.  This manifold has
$h^{1,1}=2$ and its Picard group is spanned by the restriction of the
respective hyperplanes of $\mathbb{P}^1$ and $\mathbb{P}^4$ to $X$
(respectively the line bundles $\cO(1,0)$ and $\cO(0,1)$). Now,
consider two distinct polarizations defined by $\cL_1=\cO(2,1)$ and
$\cL_2=\cO(1,2)$. We can define an embedding of $X$ into some
projective space using either of these ample line bundles. However,
the sections of each grow very differently in
$H^0(X,{\cL_{i}}^{k_{g}})$ where $i=1,2$. These sections grow with
$k_{g}$ as
\begin{equation}
  H^0\big(X,\cO(2,1)^{\otimes k_{g}}\big)
  =
  \frac{1}{3}k_{g}(23+13{k_{g}}^2)
  ,\quad
  H^0\big(X,\cO(1,2)^{\otimes k_{g}}\big)
  =
  \frac{1}{3}k_{g}(28+32{k_{g}}^2) 
  .
\end{equation}
Hence, if we computed the metric for each of these polarizations to the same degree, say $k_{g}=10$, we would have very different results. From $\cL_1$ we would have defined an embedding with $4,410$ sections, while with $\cL_2$ we would have $107,600$ sections -- and, via the algorithm of  \autoref{metric_alg}, a far more accurate approximation to the Ricci-flat metric.

In order to sensibly compare results for different polarizations then, instead of the degree of twisting $k_{g}$ (or $k_{H}$ in the case of the connection), we specify \emph{the number of sections} in $H^0(X,\cL^{k_{g}})$. When we compare the results for different polarizations, we compare them at orders chosen so that there is an approximately equal number of sections. This procedure was first introduced for Ricci flat metrics in~\cite{Braun:2007sn, Braun:2008jp}.

There is one further modification one must make to the definitions of
\autoref{review_section}. This arises in error measures used to test
the accuracy of the approximation to an Hermitian Yang-Mills
connection. In order to determine whether or not a bundle is
slope-stable for different polarizations, we must charge the
normalization of our error measure in eq.~\eqref{eq:taudef}. For
example, take the case of a general rank $n$ stable vector bundle with
structure group $U(n)$ satisfying the general Hermitian Yang-Mills
equations \eqref{hym_genr} on a $d$-dimensional K\"ahler manifold
$X$. Since $g^{i{\bar j}}F_{i{\bar j}}=\mu(\Vsheaf)\cdot \mathbf{1}_{n
  \times n}$, it is straightforward to see that in this case
\begin{equation}
  \tau(A_{\Vsheaf})=\int_X c_1(\Vsheaf \otimes \cL^{k_{H}}) \wedge \omega^{d-1} 
  ~\in \Z
  .
\end{equation}
As a result, $\tau$ manifestly depends not only on the first Chern
class of $\Vsheaf$, but also on the choice of polarization
$\omega$. In fact, it jumps by an integral amount if one changes the
polarization. In order to compare the $\tau$ error measure for
different choices of polarization, we will introduce a new
normalization that will remove the polarization dependence and make
the initial values of $\tau$ more uniform for different twists
$\Vsheaf \otimes \cL^{k_{H}}$.
\begin{figure}[tb]
  \centering
  % \framebox{
  \input{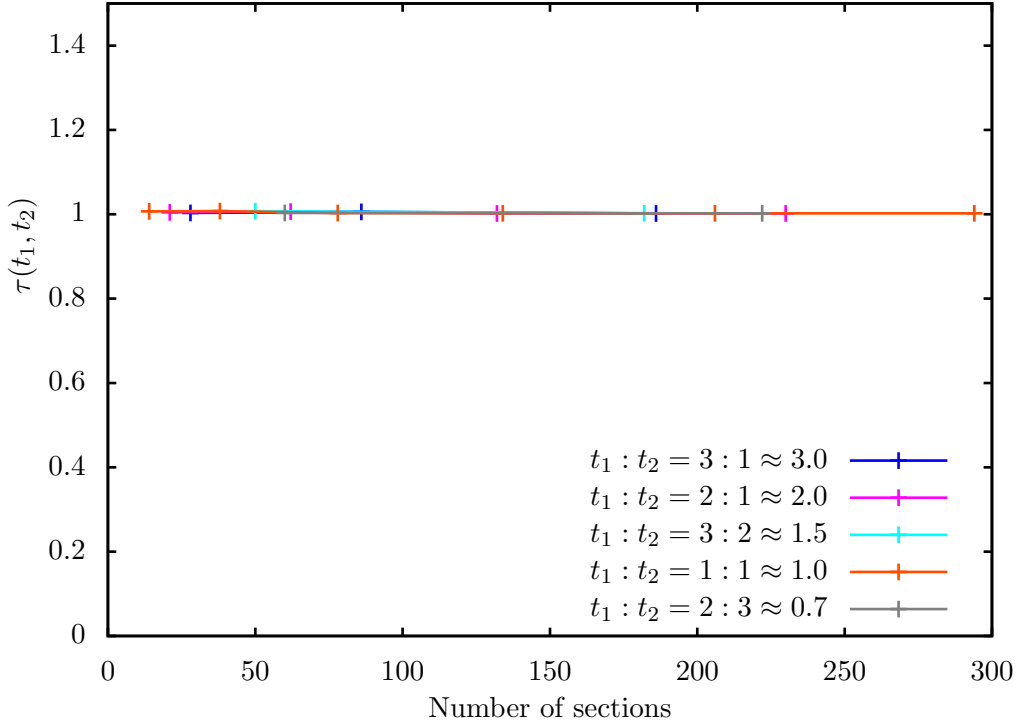}
  % }
  \caption{A plot of the normalized error measure,
    \eqref{tau_norm}. The results shown are for the sum of line
    bundles, $\Osheaf(0,1)\oplus \Osheaf(0,-1)$, on the $K3$ defined
    as a $(2,3)$ hypersurface in $\mathbb{P}^1\times
    \mathbb{P}^3$. Using the error measure introduced in
    \eqref{eq:taudef}, the results would vary for different rays in
    K\"ahler moduli space. However, with the new normalization in
    \eqref{tau_norm} the results are uniform.}
  \label{tab:Sum}
\end{figure}

As an example, take a sum of line bundles $\bigoplus_a \cL_a$ of
different first Chern class and let $\mu_{sum}=\bigoplus_a
\mu(\cL_a)$. For any given polarization
$\cL=\cO(t_r)=\cO(t_1,t_2,\ldots)$ with $r=1,\ldots, h^{1,1}$, define
$\mu_{pol}=\mu(\cL)$. We then introduce the new normalized error
measure
\begin{equation}
  \label{tau_norm}
  \tau(t_r)= \frac{\mu_{pol}}{\mu_{sum}}\tau(A_{\Vsheaf})_{k_{H}}  
  .
\end{equation}
This choice is made so that $\tau(t_r)$ will be independent of the
polarization and equal to $1$ for any sum of line bundles. This
normalization is shown in \autoref{tab:Sum} for the sum of line
bundles
\begin{equation}
  \cO(0,1)\oplus \cO(0,-1)
\end{equation}
on the $K3$ surface defined via a degree $(2,3)$ hypersurface in
$\mathbb{P}^1\times \mathbb{P}^3$. Without this normalization the
error measure of the sum of line bundles would vary according to the
polarization, making it hard to compare the different directions in
the K\"ahler moduli space. However, as is illustrated by the figure,
the error measure in \eqref{tau_norm} produces uniform results. With
these new definitions in hand, we turn to our first systematic study
of K\"ahler cone substructure.

%%%%%%%%%%%%%%%%%%%%%%%%%%%%%%%%%%%%%%%%%%%
\subsection{A Bundle On An Elliptic K3 Surface}
\label{eg_sec}

In this section, we explore the K\"ahler cone substructure described
above in an explicit example. In particular, we consider the elliptic
$K3$ surface $X$ defined as a degree $(2,3)$ hypersurface in
$\mathbb{P}^1\times \mathbb{P}^2$. This representation of the $K3$ has
$\dim (\Pic(X))=2$, that is, a $2$-dimensional algebraic K\"ahler cone,
$\Kcone$. Expanding the K\"ahler form on $X$ in a harmonic basis of
$(1,1)$-forms as $\omega=t_1 \omega_1 + t_2 \omega_2$, the K\"ahler
cone is the positive quadrant, $t_1,t_2 >0$ in the coefficients
$t_r$. Over this space, we will denote line bundles by $\cO(t_1,t_2)$,
where $\cO(1,0)$ and $\cO(0,1)$ are the pull-back of the hyperplane
bundles of $\mathbb{P}^1$ and $\mathbb{P}^2$, respectively. We will
consider below the stability properties of a sample $SU(2)$ bundle
over different polarizations in the K\"ahler cone, $\Kcone$.

We compute both the Ricci-flat metric on the base manifold and the
Hermitian Yang-Mills connection on a bundle for a number of different
polarizations. In the case of the metric, this amounts to computing
the fiber metric on the line bundle $\cL=\cO(t_1,t_2)$ for different
positive choices of $t_1,t_2$. As discussed above, in order to compare
the accuracy of these approximations to a Ricci-flat metric for
different polarizations, $t_{2}/t_{1}$, one can no longer simply
specify a twisting $\cL^{k_{g}}$. Instead, as we vary the coefficients
$t_r$, we will compare the number of global sections that are
generated by $\cL^{k_{g}}$ for different choices of $\cL$. That is,
for each choice of $\cL$, we will compute the metric up to the largest
degree $k_g$ in \autoref{metric_alg} such that there are $\leq 500$
sections in $H^0(X,\cL^{k_{g}})$. In the calculation, the metric
algorithm utilized $202,800$ points in the adaptive numeric
integration (as will be described further in \autoref{volker_int}) and
the metric T-operator was iterated $30$ times.

On this Calabi-Yau twofold we now define a rank $2$, holomorphic
vector bundle with structure group $SU(2)$. This sample bundle is
defined through the so-called monad construction~\cite{okonek,
  Anderson:2007nc, Anderson:2008uw, Anderson:2009mh},
\begin{equation}
  \label{k3_eg1}
  0 
  \longrightarrow
  \cO(-2,-1) 
  \stackrel{f}{\longrightarrow} 
  \cO(-2,0)^{\oplus 2}\oplus \cO(2,-1)
  \longrightarrow 
  \Vsheaf
  \longrightarrow
  0  .
\end{equation}
Here $\Vsheaf$ is defined as the cokernel of a generic map $f$ with
bi-degrees $((0,1), (0,1), (4,0))$ between the direct sums of line
bundles. Using the techniques of~\cite{huybrechts,Anderson:2008ex}, it
is straightforward to prove that $\Vsheaf$ in \eqref{k3_eg1} is
destabilized in part of the K\"ahler cone by the rank $1$ sheaf
\begin{equation}
\label{destabilizing}
  0 
  \longrightarrow
  \cF
\longrightarrow
  \cO(2,0)^{\oplus 2}
  \stackrel{f}{\longrightarrow} 
  \cO(2,1) 
  \longrightarrow
  0
\end{equation}
with $c_1(\cF)=(2,-1)$. The intersection numbers $d_{rs}$ of the two
hyperplane classes in the $(2,3)$ $K3$ surface are $d_{12}=d_{21}=3$,
$d_{22}=2$, and $d_{11}=0$. Using the definition of slope in
\eqref{slope}, it can be verified that $\Vsheaf$ is slope-stable when
$t_1/t_2 >4/3$ and unstable when $t_1/t_2<4/3$. That is, the K\"ahler
cone exhibits the substructure
\begin{equation}\label{eg1substruc}
  \begin{split}
    \mathcal{K}_\text{stable} 
    =&\;
    \big\{
    \Osheaf(t_1, t_2)
    \big|~
    \tfrac{t_1}{t_2} > \tfrac{4}{3}
    \big\}
    , \\
    \mathcal{K}_\text{unstable} 
    =&\;
    \big\{
    \Osheaf(t_1, t_2)
    \big|~
    \tfrac{t_1}{t_2} < \tfrac{4}{3} 
    \big\}  .
  \end{split}
\end{equation}
Hence, in the stable region we expect the T-operator to converge,
whereas in the unstable region we do not expect to find a fixed point.

As discussed in \autoref{connection_alg}, to apply the generalized
Donaldson algorithm one must define the embedding $i_k: X \to
G(n,N_{k_{H}})$. To do this, one must compute the global sections of
the twisted line bundle $\Vsheaf \otimes \cL^{k_H}$ for some ample
line bundle $\cL$. For the bundle defined in \eqref{k3_eg1}, the
global sections $H^0(X,\Vsheaf \otimes \cL^{k_H})$ can be computed for
any choice of twisting. Multiplying eq.~\eqref{k3_eg1} by
$\cL^{k_H}=\cO(t_1,t_2)^{k_{H}}$, we obtain the short exact sequence
\begin{multline}
{\tiny  0 
  \to
  \cO(k_{H}t_1-2,k_{H}t_2-1) 
  \stackrel{f}{\longrightarrow}
  \cO(k_{H}t_1-2,k_{H}t_2)^{\oplus 2} \oplus 
  \cO(k_{H}t_1+2,k_{H}t_2-1)} \\
 {\tiny \to
  \Vsheaf \otimes \cO(t_1,t_2)^{\otimes k_{H}} 
  \to
  0}.
  \label{aaa}
\end{multline}
\begin{figure}[tb]
  \centering
 % \framebox{
    \input{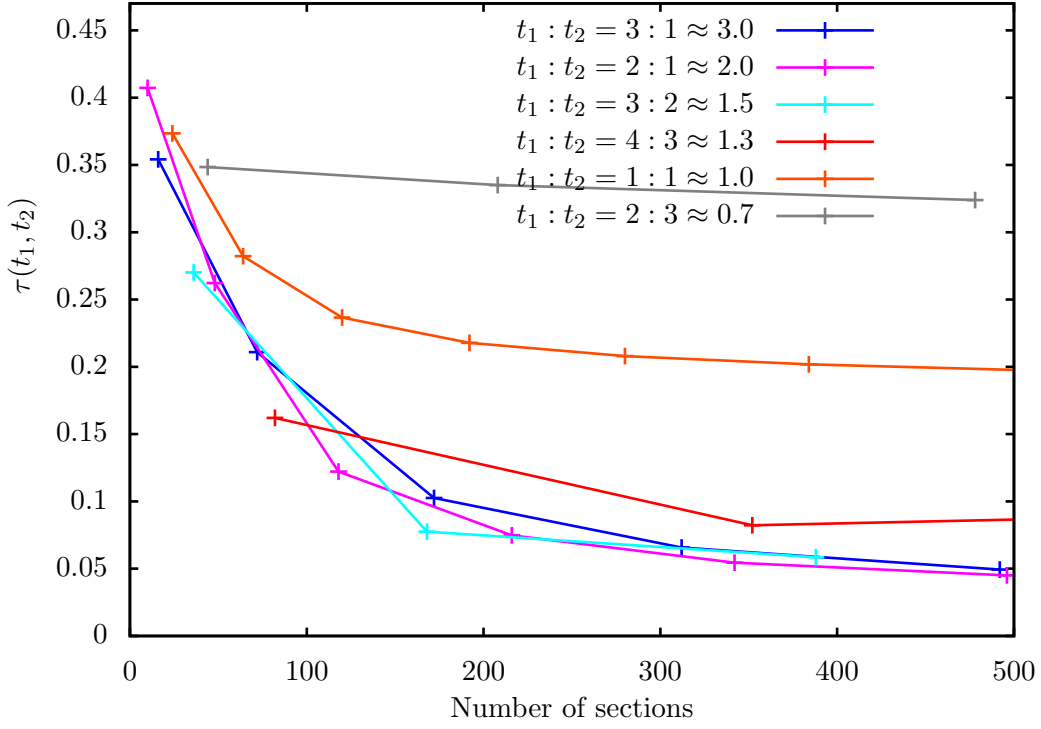}
%  }
  \caption{The K\"ahler cone substructure associated with the $SU(2)$ bundle in \eqref{k3_eg1}. The normalized error measure, \eqref{tau_norm}, is shown for different choices of polarization in the K\"ahler cone. The presence of $\Vsheaf$ clearly divides the K\"ahler cone, $\Kcone$, into two chambers, $\Kcone_\text{stable}$ and $\Kcone_\text{unstable}$, corresponding to the stable/unstable regions described in \eqref{eg1substruc}.}
  \label{tab:Substructure}
\end{figure}
Then the global sections are given simply as the cokernel
\begin{equation}\label{h0_def}
  H^0\big(X,\Vsheaf \otimes \cO(t_1,t_2)^{k_{H}} \big)
  = 
  \frac{
    H^0\big(X,
    \cO(k_{H}t_1-2,k_{H}t_2)^{\oplus 2}\oplus
    \cO(k_{H}t_1+2,k_{H}t_2-1)
    \big)
  }{
    f\big(H^0(X, \cO(k_{H}t_1-2,k_{H}t_2-1))\big)
  }  
  ,
\end{equation}
where both parts of this quotient are the global sections of
sums of ample line bundles when $k_{H}t_1> 2$ and $k_{H}t_2>1$. 

For $\Vsheaf$ in \eqref{k3_eg1}, we will apply the generalized
Donaldson algorithm as reviewed in \autoref{connection_alg} and
\autoref{algorithm_review}. In particular, we will compute Hermitian
bundle metrics on $\Vsheaf$ and determine whether they converge to the
Hermitian Yang-Mills connection for different choices of
polarization. Specifically, we will compare the results for six
different choices of polarization: three lying within
$\Kcone_\text{stable}$, two in $\Kcone_\text{unstable}$ and one on the
boundary defined by $t_1/t_2=4/3$.  The results are shown in
\autoref{tab:Substructure}, where we plot the normalized $L^1$ error
measure, eq.~\eqref{tau_norm}, for the Yang-Mills connection in
various directions in the K\"ahler cone. Since $\Vsheaf$ is a cokernel
associated with the direct sum $\bigoplus_i \cL_i=\cO(-2,0)^{\oplus
  2}\oplus \cO(2,-1)$, the normalization $\mu_{sum}$ in
eq.~\eqref{tau_norm} was chosen to be
$\mu_{sum}=2\mu\big(\cO(-2,0))+\mu(\cO(2,-1)\big)$ in order to
meaningfully compare the different rays in the K\"ahler cone. In the
stable sub-cone, the eigenvalues of $g^{i{\bar j}}F_{i{\bar j}}$ can
clearly be seen to be approaching zero as one increases the number of
sections in $H^0(X,\Vsheaf\otimes \cL^{k_H})$, as expected. In the
unstable region there is no such convergence. Both observations are in
perfect agreement with eq.~\eqref{eg1substruc}. The connection on
$\Vsheaf$ was computed with $N_G = 74,892$ points (adaptive) and the
connection T-operator was computed with $100$ iterations at each graph
point in \autoref{tab:Substructure}.

The results of this section clearly indicate that the generalized
Donaldson algorithm can be used to investigate K\"ahler cone
sub-structure. We will explore this in more detail in the following
sections, but first we will provide a description of the novel
integration method implemented and used throughout this work. This
integration scheme provides a significant increase in computation
speed and makes it possible for us to analyze a wider range of
examples.

\section{Adaptive Integration}\label{volker_int}

\subsection{Rectangle Method vs. Monte-Carlo}

All approaches to numerical geometry of Calabi-Yau threefolds, be it
Donaldson's algorithm~\cite{MR2161248, MR1916953, DonaldsonNumerical,MR2283416, Braun:2007sn, Braun:2008jp, Douglas:2006rr,
  Douglas:2006hz, Headrick:2009jz,Anderson:2010ke}, its generalization to Hermitian
Yang-Mills bundles~\cite{MR2154820,Douglas:2006hz,Anderson:2010ke}, or direct
minimization~\cite{Headrick:2009jz} all use a spectral representation
of the geometric data. That is, the tensors describing the geometry
are eventually expanded in a suitable basis of functions, and the
problem reduces to finding the ``best fit'' coefficients. This is in
contrast to the traditional finite elements methods, where one
directly discretizes spacetime. As a general rule, finite elements
work well in low dimensions, but spectral representations are
necessary in high-dimensional problems.

The basic numerical step that every spectral algorithm relies on at
its core is to integrate over the base manifold. This is necessarily
so because only by evaluating the spectral basis \emph{everywhere} on
the manifold can one draw conclusions about the global behavior of the
geometric object of interest. The most straightforward integration
scheme is to split the integration domain into equal-sized pieces and
use the multidimensional generalization of the rectangle rule. In
practice, the volume elements can only be chosen of equal size with
respect to an auxiliary metric. But this is then easily corrected for
by weighting the individual points accordingly,
\begin{equation}
  \sqrt{g} \; d^n x =
  w(x) \; \sqrt{g_\text{aux}} \; d^n x 
  \quad \Rightarrow \quad
  w(x) 
  = \frac{\sqrt{g}}{\sqrt{g_\text{aux}}}
  .
\end{equation}
Here, the scalar weight function $w(x)$ is simply determined by the
auxiliary measure (given by the point distribution) and the desired
measure $\sqrt{g}\; d^n x$. The Calabi-Yau case is particularly
simple, since the Calabi-Yau volume form $\sqrt{g_\text{CY}}\; d^n x =
\Omega\wedge \overline \Omega$ is known analytically.

The disadvantage of this direct approach is that it requires extensive
knowledge about the geometry and topology of the manifold to construct
a constant (auxiliary) volume cell decomposition. This is why it has
only been used for complex surfaces~\cite{2005math.....12625D,
  2008arXiv0803.0987B}, that is, real 4-dimensional manifolds. A
solution to this problem was devised in~\cite{Douglas:2006rr}, where a
Monte-Carlo integration scheme was proposed that needs as input only
the defining equation of a Calabi-Yau hypersurface (or complete
intersection). The key to this approach is that one can generate
random points with known distributions using zeroes of random sections
of line bundles~\cite{Zelditch:Shif, MR1616718}. As an example,
consider the quintic $i:Q\to\CP^4$ embedded in projective space. A
line $\CP^1\subset \CP^4$ is defined by three linear equations in the
homogeneous coordinates, that is, three sections of
$\Osheaf_{\CP^4}(1)$. Any one linear equation is defined by its $5$
coefficients, so one can talk about random sections with an
$SU(5)$-invariant distribution. Three linear equations intersect the
quintic hypersurface in $5$ points, and by the general theory the
probability distribution has measure $i^*(\omega_\text{FS})^3$. The
Calabi-Yau volume form is given by the residue integral
\begin{equation}
  \Omega = \oint \frac{d^4 \rho}{Q(\rho)}
\end{equation}
in a local (holomorphic) coordinate patch $(\rho_1,\rho_2,\rho_3,\rho_4)$, thus
determining the weight function $w(x)$ for the random points. The
disadvantage of this Monte-Carlo integration scheme is that one is
forced to use the random point set without modification. In
particular, it tuns out that the point weights $w(x)$ fluctuate over a
large range with the error accumulating in the badly-sampled
regions. This was mitigated in\cite{Keller:2009vj} using stratified
sampling, at the cost of having to work with higher-dimensional
Kodaira embeddings.

\subsection{An Adaptive Integration Algorithm}

For the purposes of this paper, we developed a combination of the best
features of the naive higher-dimensional rectangle rule and the
Monte-Carlo sampling. In a nutshell, the idea is to first parametrize
point(s) of the Calabi-Yau threefold similarly to the parametrization
by sections of three line bundles. Then integrate using the standard
higher-dimensional rectangle rule by constructing a suitable cell
decomposition of the \emph{parameter space}, not the Calabi-Yau
manifold. 

As the simplest example, consider the $(2,3)$-hypersurface in
$X\subset \CP^1\times\CP^2$, which is a $K3$ surface. The projection
$\pi:X\to\CP^2$ onto the $\CP^2$ factor is, generically,
two-to-one. Locally, there are two inverses $\pi^{-1}_1$, $\pi^{-1}_2$
for the two-sheeted cover $\pi$. With it, one can rewrite the
integration as
\begin{equation}
  \int_X f(x) \sqrt{g} \; d^4x 
  =
  \int_{\CP^2} 
  \sum_{i=1}^2
  f\big(\pi^{-1}_i(z)\big) 
  \underbrace{
    \left|\frac{\partial\pi^{-1}_i}{\partial z}\right|
    \;
    \frac{\sqrt{g}}{\sqrt{g_\text{aux}}}
  }_{w_i(z)}
  \;
  \sqrt{g_\text{aux}} \; d^4z  ,
\end{equation}
where the sum runs over the different sheets. The weights $w_i(z)$ are
the product of the Jacobian of the coordinate transformation and, as
before, the scalar factor required to transform the auxiliary measure
into the desired Calabi-Yau measure. Finally, it is easy to integrate
over projective space. For our purposes, we will use an adaptive
integration scheme where we start with a decomposition of $\CP^2$ into
cells of equal volume with respect to some convenient auxiliary
measure. Then, if any of the weights $w_i(z)$ is significantly larger
than the average weight, we recursively subdivide the cell until the
weight is acceptably small. Finally, we use the usual rectangle rule
and sum over all cells to compute the integral.

It is not really necessary for the map $\pi:X\to \CP^2$ to be a
multisheeted cover or even to be defined everywhere. In particular, a
dominant rational map would be perfectly fine. To summarize, our
integration algorithm
\begin{itemize}
\item does not require a cell decomposition of the Calabi-Yau manifold,
\item converges as $O(\tfrac{1}{N})$ with the number of
  points, just like the standard rectangle rule, and
\item makes it easy to adapt each integration step to reduce numerical
  errors.
\end{itemize}

\subsection{Integrating Over Projective Space}

Thus far, we reduced the integration to one over projective space
$\CP^n$ with some convenient auxiliary metric. We now describe a way
of decomposing $\CP^n$ into equal-sized cells that is suitable for
adaptive subdivision of the cells. First, let us decompose $\CP^n$
into $n+1$ polydiscs
\begin{equation}
  \CP^n = \bigcup_{i=0}^n D_i
  ,\qquad
  D_i = 
  \Big\{
  [z_0:
  \cdots:z_{i-1}:
  \underbrace{1}_{\mathclap{\text{$i$-th entry}}}:
  z_{i+1}:\cdots:
  z_n]
  ~
  \Big|
  ~
  |z_j|\leq 1
  \Big\}
  \simeq 
  D^n
  .
\end{equation}
%%%%%%%%%%%%%%%%%%%%%%%%%%%%%%%%%%%%%%%%%%
\begin{figure}[p]
  \centering
  \begin{tabular}{c@{\ }c}
    & \parbox{12cm}{\hfill 
      coarse ($\approx 20$ points)\hfill\hfill
      medium ($\approx 80$ points)\hfill\hfill 
      fine ($\approx 300$ points)\hfill \strut}\\
    \begin{sideways}
      \parbox{4.28cm}{\centering Euclidean: $dz\wedge d\bar{z}$}
    \end{sideways}
    & \includegraphics{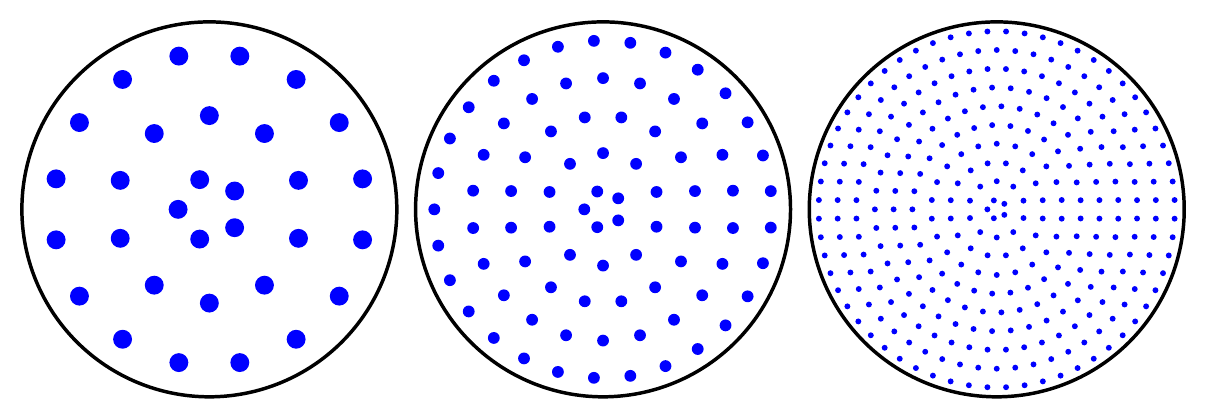} \\
    \begin{sideways}
      \parbox{4.28cm}{\centering Fubini-Study: $\omega_{FS}$}
    \end{sideways}
    & \includegraphics{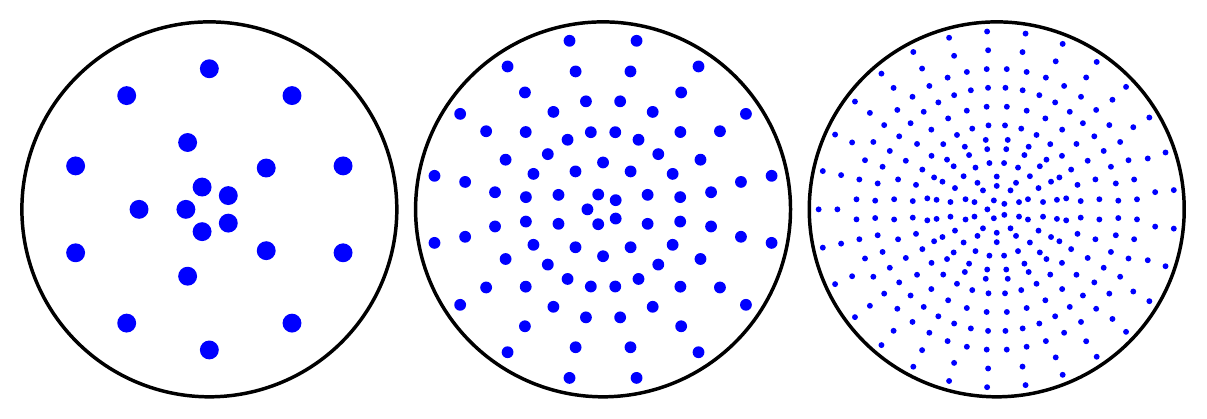} \\
    \begin{sideways}
      \parbox{4.28cm}{\centering Calabi-Yau: $\Omega\wedge\overline{\Omega}$}
    \end{sideways}
    & \includegraphics{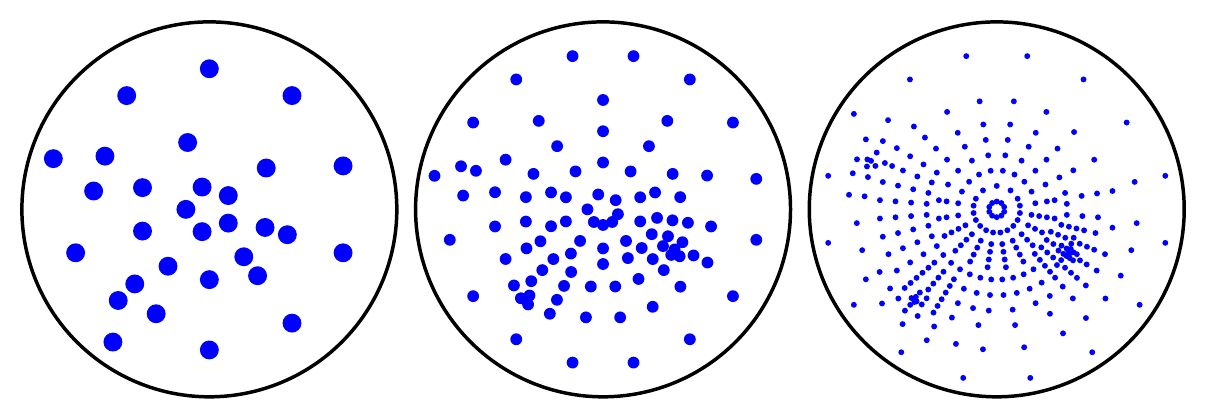} 
  \end{tabular}
  \caption{The adaptive mesh generation for four different
    integrations over the disk $D=\{|z|\leq 1\}$. 
    In each case, the points are chosen such that their weight is
    approximately constant. The point distributions in the three rows
    are further explained in items~\ref{item:adapt1},
    \ref{item:adapt2}, and \ref{item:adapt3} on page~\pageref{item:adapt1}.}
  \label{fig:adapt}
\end{figure}
Note that, by dividing with the homogeneous coordinate of largest
magnitude, the homogeneous coordinates of any point can be rescaled
such that
\begin{itemize}
\item one homogeneous coordinate equals unity, say, $z_i=1$, and 
\item all other homogeneous coordinates have equal or smaller
  magnitude $|z_j|\leq 1$, $j\not=i$.
\end{itemize}
Hence, a generic point of $\CP^n$ is contained in precisely one of the
polydiscs $D_i$. The polydiscs overlap in the measure-zero set where
two or more homogeneous coordinates attain the maximum magnitude.

It remains to decompose each polydisc $D_i\simeq D^n$. For this
purpose, we use that the polydisc is the Cartesian product of $n$
individual disks
\begin{equation}
  D = \Big\{
  r e^{i\varphi}
  ~\Big|~ 
  r\leq 1 
  ,\;
  0\leq \varphi <2\pi
  \Big\} \subset \C
  .
\end{equation}
Any cell decomposition of $D\subset \C$ induces one of the
polydisc. For our implementation, we chose to decompose $D$ into
annuli of width $\delta r$, and each annulus into segments of angles
$\delta \varphi \approx \delta r /(2\pi r)$. The annulus segments are
approximately quadratic and of area $2\pi r \delta \varphi \cdot
\delta r$ in the Euclidean measure. 

To summarize, we decompose the polydiscs $D_i$ into hypercubes of
constant volume in the Euclidean metric, which we use as the auxiliary
measure. If the weight of the cell is too large, we subdivide it by
splitting the annulus segment in one of the $n$ discs\footnote{We
  chose one of the $n$ discs randomly, but this could clearly be
  improved by separating the cell in the direction of the biggest
  gradient for the weights.}  that constitute $D_i$. Summing over each
of the $n+1$ polydiscs according to the rectangle rule then computes
the integral over $\CP^n$. To illustrate the adaptive integration
algorithm, we plot the point distribution for three different volume
forms in \autoref{fig:adapt}. The first, second, and third column
shows successive refinements with growing number of points. Each row
corresponds to a different volume form that the point distribution is
adapted to:
\begin{enumerate}
\item\label{item:adapt1} In the top row, points are distributed
  regularly on the disc. This distribution, with constant weight
  attached to each point, approximates integration with respect to the
  Euclidean measure.
\item\label{item:adapt2} In the second row, we try to approximate
  integration with respect to the Fubini-Study measure. Here, the
  points are chosen adaptively and are denser towards the center where
  the Fubini-Study volume form is denser.
\item\label{item:adapt3} In the third row, we illustrate integration
  over the $(2,2)$-hypersurface 
  \begin{multline}
    w_0^2 z_0^2
    - \tfrac{9+i}{10} w_0^2 z_0 z_1 
    - \tfrac{3-9i}{10} w_0^2 z_1^2
    - \tfrac{1-7i}{10} w_0 w_1 z_0^2
    - \tfrac{5-10i}{10} w_0 w_1 z_0 z_1
    \\
    - \tfrac{4-i}{10} w_0 w_1 z_1^2
    - \tfrac{10+9i}{10} w_1^2 z_0^2 
    - \tfrac{2+2i}{10} w_1^2 z_0 z_1
    + \tfrac{5-i}{10} w_1^2 z_1^2
    = 0
  \end{multline}
  in $\CP^1_{[w_0:w_1]}\times\CP^1_{[z_0:z_1]}$ with respect to the
  Calabi-Yau volume form. Using the projection
  $\pi:X\to\CP^1_{[z_0:z_1]}$ we rewrite this integration as the
  integration over a single $\CP^1$. The disc in \autoref{fig:adapt}
  shows the first patch $[z_0:z_1] = [z:1]$, $|z|\leq 1$. The points
  are adaptively chosen to have approximately constant weight.
\end{enumerate}

%%%%%%%%%%%%%%%%%%%%%%%%%%%%%%%%%%%%%%%
\section{Diagnosing Stability}\label{t_iter_error}

One of the goals of this paper to use the generalized Donaldson
algorithm to probe K\"ahler cone substructure in higher-dimensional
K\"ahler cones. That is, we would like to efficiently be able to scan
a K\"ahler cone for regions where a given bundle is stable. To this
end, we present in this section a new and efficient numerical measure
of the stability properties of $\Vsheaf$ at a fixed polarization.

The central idea is as follows. If one just wants to numerically
determine whether a bundle is stable or not, it is not necessary to
explicitly compute the Hermitian Yang-Mills connection. Instead, one
can simply use the first step of the generalized Donaldson algorithm,
namely the convergence properties of the T-operator.  As we saw in
\autoref{connection_alg}, according to Wang's theorem, for a fixed
embedding defined by $H^0(X, \Vsheaf\otimes \cL^{k_{H}})$, the
T-operator can be moved to a balanced place if and only if it is
Gieseker stable as defined in eq.~\eqref{gieseker}. Thus, for an
embedding defined by even a relatively small number of sections (that
is, a small degree of twisting $k_H$), it should be straightforward to
check whether or not the iterations of the T-operator in \eqref{t_gen}
converge to a fixed point. If such a fixed point exists, then we know
that the bundle is Giesker stable. In general, this computation is
easier and faster than a complete computation of the connection and
its integrated error measure.
\begin{figure}[htb]
  \centering
 % \framebox{
    \input{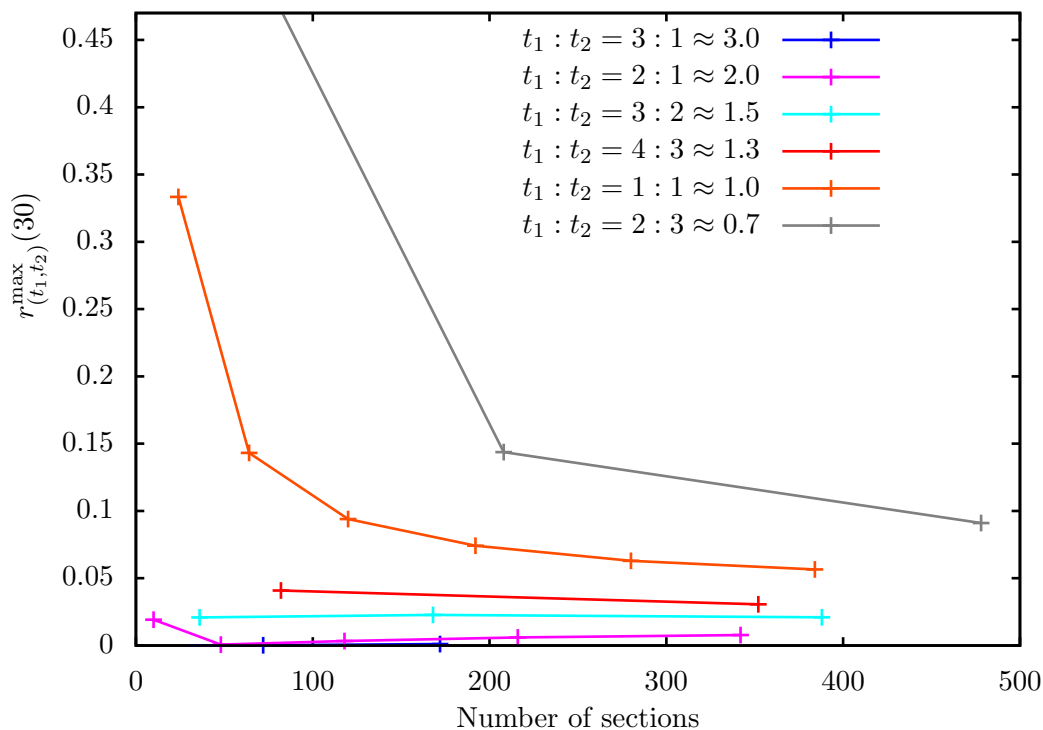}
%  }
  \caption{The T-operator convergence measured for the same bundle
    and polarizations as in \autoref{tab:Substructure}.}
  \label{tab:r}
\end{figure}

As part of our program of probing the stability of general bundles
$\Vsheaf$ in higher dimensional K\"ahler cones, it is intriguing to
see that one can gain important information regarding Gieseker
stability from the T-operator and only a minimal number of twists
$\Vsheaf \otimes \cL^{k_{H}}$. However, one must be careful. While it
is certainly true that all slope-stable bundles are Gieseker stable,
properly Gieseker stable bundles need only be slope semi-stable. We
can only conclude that if the T-operator fails to converge, then
$\Vsheaf$ is \emph{not} slope-stable. However, if it does converge, we
only know that $\Vsheaf$ is semi-stable, but not necessarily a
solution to \eqref{hym_genr}. In order to be certain that $\Vsheaf$ is properly slope-stable, one would then need to augment the analysis of the T-iteration with a full computation of the $\tau$ error measure in \eqref{tau_norm}, as discussed in the previous sections. In the next section, we
will investigate this subtle semi-stable behavior in detail. For now, however, we
only explore how much can be gained from a simple check of T-operator
convergence.

With this idea in mind, we develop a new error measure based on the convergence of the T-operator. Considering the matrix $H_{\alpha\bar\beta}$ in \eqref{T_gen_bal}, we would like to know how much the matrix changes as one moves from the $m$-th to the $(m+1)$-th iteration of the T-operator in \eqref{t_gen}. To answer this question, consider the eigenvalues of the $H_{\alpha\bar\beta}$-matrix. At each step of the
iteration, the largest eigenvalues are the relevant features; small
eigenvalues only give small corrections to the connection on the
bundle. Let
\begin{equation}
  v_{t_1,t_2}^\text{max}(m)
\end{equation}
be the largest eigenvalue of the $H_{\alpha\bar\beta}$ after $m$
iterations of the T-operator. Except for the overall numerical scale, one expects
that the details of the initial values are washed out by the
iteration. However, the overall scale is preserved by the iteration
and, moreover, does not enter the connection. Therefore, a good
quantity to measure the convergence is 
\begin{equation}\label{r_def}
  r_{(t_1,t_2)}^\text{max}(m) = 
  \frac{
    v_{(t_1,t_2)}^\text{max}(m)
  }{
    v_{(t_1,t_2)}^\text{max}(m-1)
  }
  - 1
  .
\end{equation}
By \autoref{wang2} and the above considerations, we know that
\begin{equation}
  \lim_{m\to \infty} r_{(t_1,t_2)}^\text{max}(m) =0  
\end{equation}
if $\Vsheaf$ is Gieseker stable as in eq.~\eqref{gieseker}. For the
purposes of this paper, we are interested in slope-stability and
solutions to the Hermitian Yang-Mills equations. Since slope-stability
implies Gieseker stability, it follows that
$r_{(t_1,t_2)}^\text{max}(m)$ should have the following behavior in
the presence of K\"ahler cone substructure:
\begin{equation}
  \lim_{m\to \infty}
  r_{(t_1,t_2)}^\text{max}(m) 
  \quad
  \begin{cases}
    \quad
    = 0 & \text{if}~ 
    (t_1,t_2) \in \mathcal{K}_\text{stable}
    \\
    \quad
    > 0 & \text{if}~ 
    (t_1,t_2) \in \mathcal{K}_\text{unstable}
  \end{cases}
\end{equation}

We will put this to the test in the example introduced in \autoref{eg_sec}. For the $SU(2)$ 
bundle $\Vsheaf$ in \eqref{k3_eg1}, one would expect
\begin{equation}
  \lim_{m\to \infty}
  r_{(t_1,t_2)}^\text{max}(m) 
  \quad
  \begin{cases}
    \quad
    = 0 & \text{if}~ 
   \frac{t_1}{t_2}> \frac{4}{3}
    \\
    \quad
    > 0 & \text{if}~ 
   \frac{t_1}{t_2} < \frac{4}{3}
  \end{cases}
\end{equation}
The results are shown in \autoref{tab:r} for the same six polarizations chosen in \autoref{tab:Substructure}. The connection T-operator was computed at $30$ iterations and the comparison of \eqref{r_def} made between the $29$th and $30$th iterations. As expected, $r_{(t_1,t_2)}^\text{max}(30)$ is approximately zero in the slope-stable region of K\"ahler moduli space, but non-zero in the unstable region.

\begin{figure}[htb]
  \centering
  %\framebox{
  \input{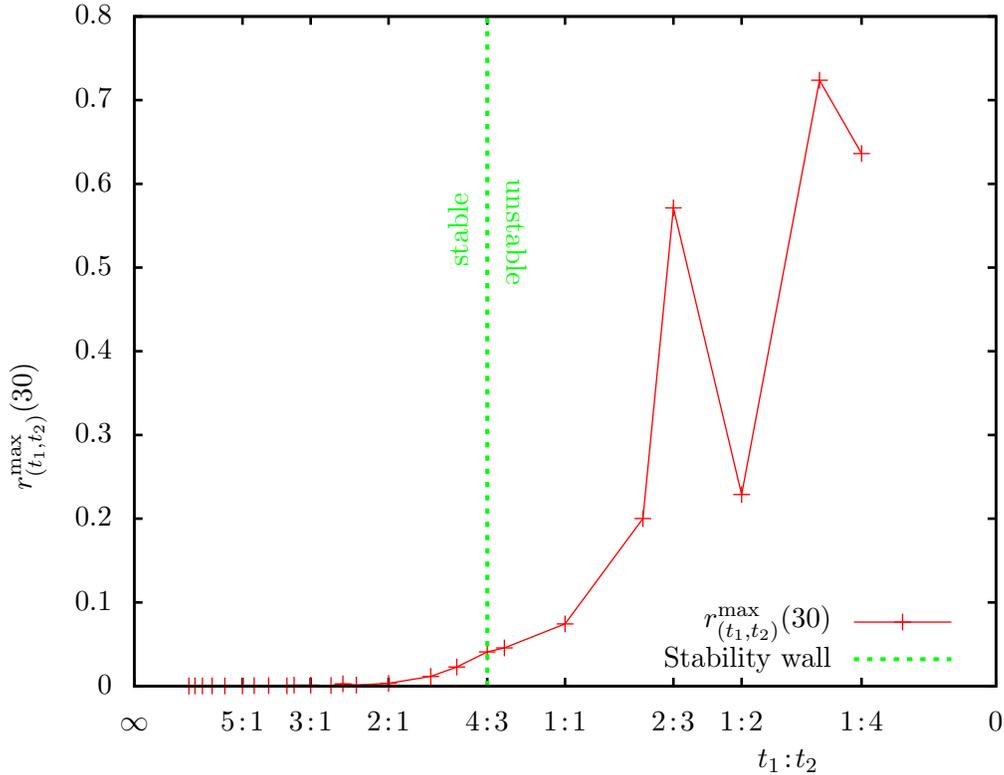}
  %}
  \caption{Convergence of the T-iteration on the $SU(2)$ bundle in \eqref{k3_eg1} with
    substructure for different K\"ahler moduli. The
    radial direction $\gcd(t_1,t_2)$ is always chosen to be as large
    as possible subject to the constraint that there are $\dim
    H^0\big( \Osheaf(t_1,t_2) \big) \leq 200$ sections.}
  \label{tab:Angular}
\end{figure}

On the boundary between $\Kcone_\text{stable}$ and
$\Kcone_\text{unstable}$, however, we find that the
$r_{(4,3)}^\text{max}(30)$ is also close to zero, despite the fact
that $\Vsheaf$ in \eqref{k3_eg1} is only slope semi-stable for this
polarization. This is to be expected however, since it can be shown
via direct computation (and \eqref{gieseker}) that $\Vsheaf$ is still
properly Gieseker stable for this line in K\"ahler moduli space and
Gieseker stability implies only slope semi-stability. It follows that
to accurately determine the behavior on this boundary, one would need
to also investigate the $\tau$ error measure of the previous section,
which can distinguish between slope-stable and semi-stable
behavior. We return to the boundary behavior in the next section.

\begin{figure}[tbp]
  \centering
  %\framebox{
    %\includegraphics[width=10cm,
    %height=10cm]{StabilityWall-EllipticK3-2D.png}
    \input{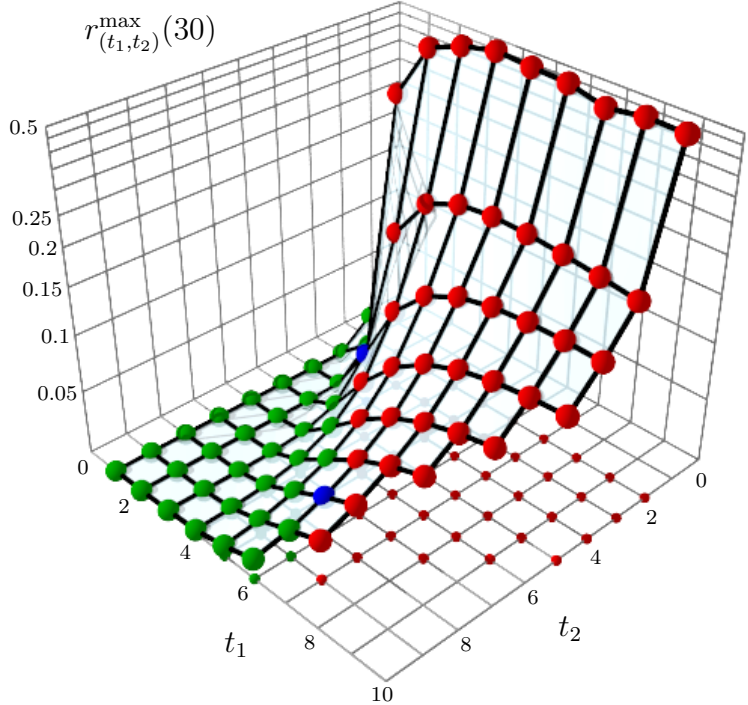}
  %}
  \caption{The K\"ahler cone substructure associated with the $SU(2)$ bundle in \eqref{k3_eg1} for different
    K\"ahler moduli. Green points are stable, blue points are on the
    line of stability, and red points are unstable polarizations.}
  \label{tab:Substructure2D}
\end{figure}

For now, it should be noted that one need not have checked the
T-operator for all $500$ sections in \autoref{tab:r}. In fact, the
convergence of the T-operator is already evident at a much smaller
projective embedding. While we must have enough sections to make sure
that \eqref{bundle_embed} is a proper embedding, we can see the
stability properties of $\Vsheaf$ from the first embedding which makes
$H^0(X,\Vsheaf \otimes \cL^{k_{H}})$ non-vanishing. In
Figures~\ref{tab:Angular} and~\ref{tab:Substructure2D} we present the
same results from a purely ``angular'' point of view; that is,
computing the T-operator at only a single point along each of the rays
plotted in \autoref{tab:r}. These points (that is, the twistings
$\cL^{k_{H}}$) were chosen so that there are $ \leq 200$ sections at
each point. Note that the oscillation in the height of the points in
\autoref{tab:Angular} in the unstable region is not significant since
we are comparing data from different polarizations, which leads to
different normalizations. The only meaningful comparison is between
zero and non-zero values of $r_{(t_1,t_2)}^{max}(m)$. We also present
a $3$-dimensional plot of the same K\"ahler cone substructure in
\autoref{tab:Substructure2D}. For all calculations, the connection was
computed with $N_G = 74,892$ points (adaptive) and $30$ iterations of
the T-operator.

\begin{figure}[htb]
  \centering
 % \framebox{
  \input{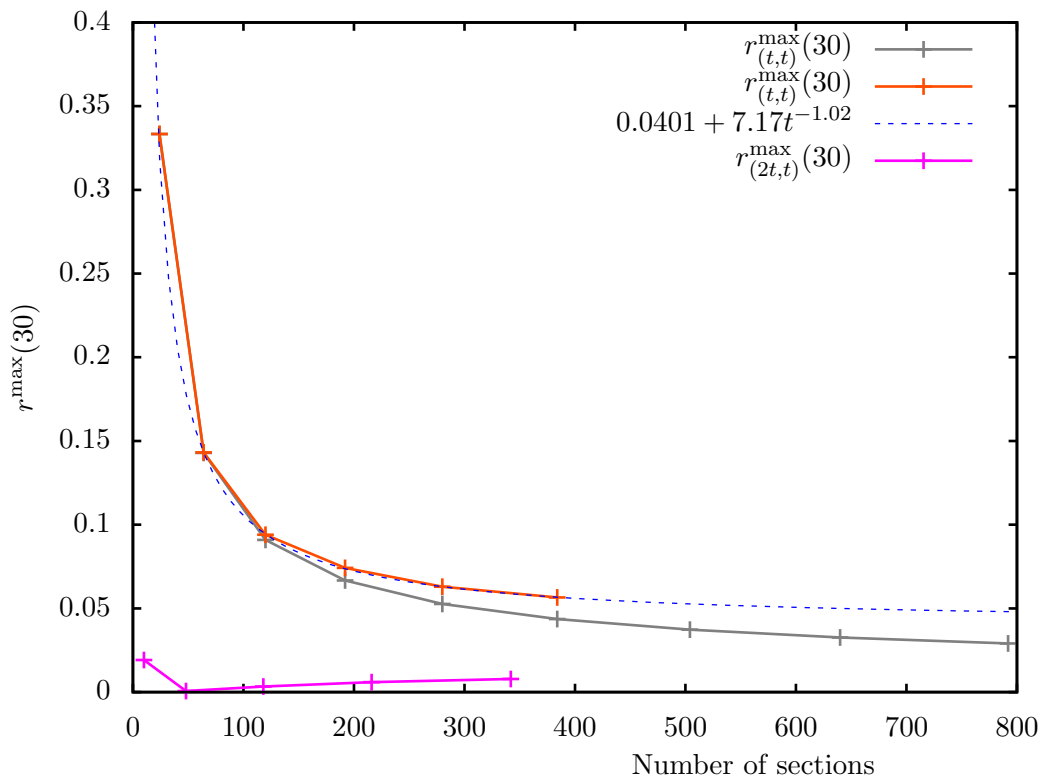}
%  }
  \caption{Convergence (for the polarization $\Osheaf(2t,t)$) and
    divergence (for $\Osheaf(t,t)$) of the T-iteration on the bundle
    $\Vsheaf$ in \eqref{k3_eg1}. Substructure is shown for different K\"ahler moduli
    along a stable and an unstable ray, respectively. The orange and gray lines are associated with two different values of the bundle moduli for the unstable polarization.}
  \label{tab:ev-k}
\end{figure}

Finally, in \autoref{tab:ev-k} we consider the rate at which the
T-operator approaches its fixed point as the number of sections
defining the embedding is increased, see eq.~\eqref{bundle_embed}. We
observe that convergence of the T-iteration is generally slower for
more sections. In particular, in \autoref{tab:ev-k} we present the
convergence (and divergence) of the T-iteration for polarizations
$\Osheaf(t,t)$, which is in the unstable region
$\Kcone_\text{unstable}$ of the K\"ahler moduli space.

\section{On The Line Of Semi-Stability}\label{stab_wall_sec}

In the previous sections, we demonstrated that the generalized
Donaldson algorithm is capable of broadly probing K\"ahler cone
substructure. In this section, we take a detailed look at the boundary
between stable/unstable regions, the so-called ``stability
wall"~\cite{Anderson:2009sw,Anderson:2009nt}.

\begin{figure}[htb]
  \centering
 % \framebox{
  \input{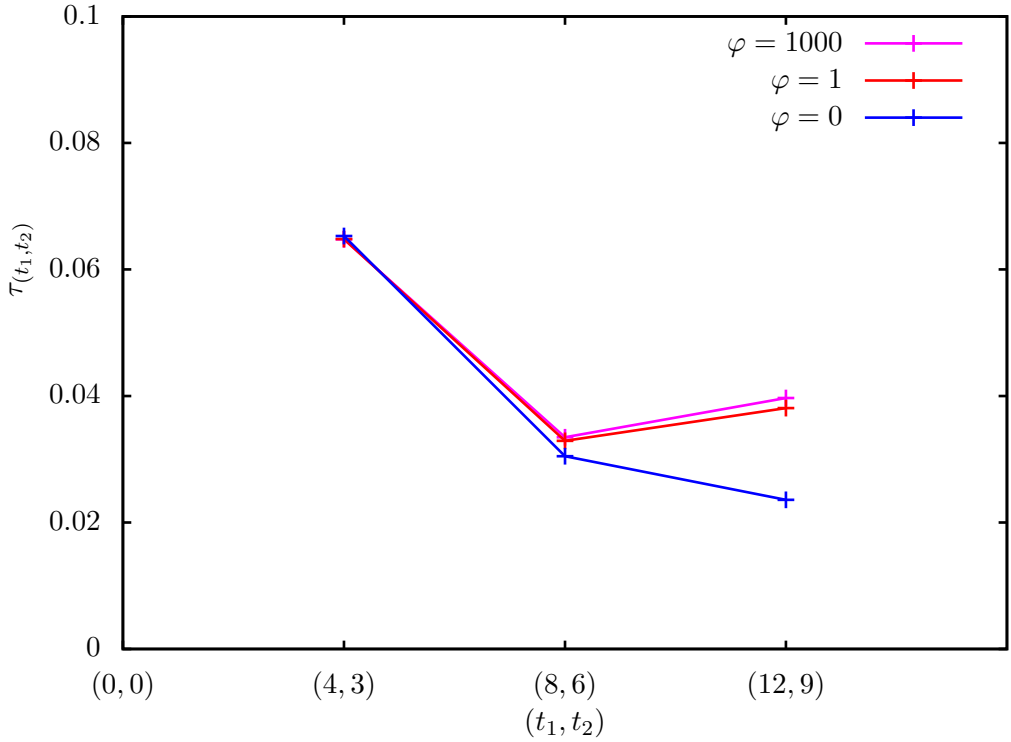}
%  }
  \caption{The $\tau$ error measure associated with the $SU(2)$ bundle in \eqref{k3_eg1} at the stability wall in K\"ahler moduli space. The bundle modulus $\varphi$ defined in \eqref{f_varphi} determines whether $\Vsheaf$ is strictly semi-stable (when $\varphi \neq 0$) or poly-stable (when $\varphi=0$). In the latter case, the reducible connection satisfies the Hermitian Yang-Mills equations \eqref{the_hym}.
}
  \label{tab:Wall}
\end{figure}

Let us revisit the $SU(2)$ bundle $\Vsheaf$ on the $K3$ surface
defined by \eqref{k3_eg1}. As discussed in \autoref{eg_sec}, $\Vsheaf$
is destabilized in part of the K\"ahler cone by a rank $1$ subsheaf
$\cF \subset \Vsheaf$ defined in \eqref{destabilizing}. In the region
$\Kcone_\text{stable}$ in \eqref{eg1substruc}, $\mu(\cF)<0$ and in
$\Kcone_\text{unstable}$, $\mu(\cF)>0$. What happens on the boundary
line between these two regions, where $\mu(\cF)=\mu(\Vsheaf)=0$?  By
definition, on a line with $\mu(\cF)=\mu(\Vsheaf)$ the bundle
$\Vsheaf$ is semi-stable. Hence, for generic values of the bundle
moduli its connection will not solve the Hermitian Yang-Mills
equations.  However, semi-stable bundles are distinguished from
unstable bundles in that they can provide supersymmetric solutions for
special loci in their moduli space. Looking at the definitions in
\eqref{slope_req} and \eqref{slope_themovie} in
\autoref{review_section}, we see that the only way that $\Vsheaf$ can
satisfy the Hermitian Yang-Mills equations when
$\mu(\cF)=\mu(\Vsheaf)$ is for it to be \emph{poly-stable} rather than
strictly semi-stable. That is, if the connection on $\Vsheaf$ is
decomposable, it is possible that a connection may exist which
satisfies \eqref{hym_genr}. Mathematically, this property of slope
semi-stable bundles is characterized by the notion of S-equivalence
classes and the Harder-Narasimhan filtration \cite{huybrechts,
  MR2180559}, which states that every semi-stable bundle has a unique
poly-stable representative in its moduli space.

For the bundle in \eqref{k3_eg1}, we find that the poly-stable representative arises when $\Vsheaf$ decomposes on the semi-stable wall as the direct sum
\begin{equation}\label{split}
  \Vsheaf \longrightarrow \cF \oplus \cO(-2,1)
  .
\end{equation}
\begin{figure}[htb]
  \centering
 % \framebox{
  \input{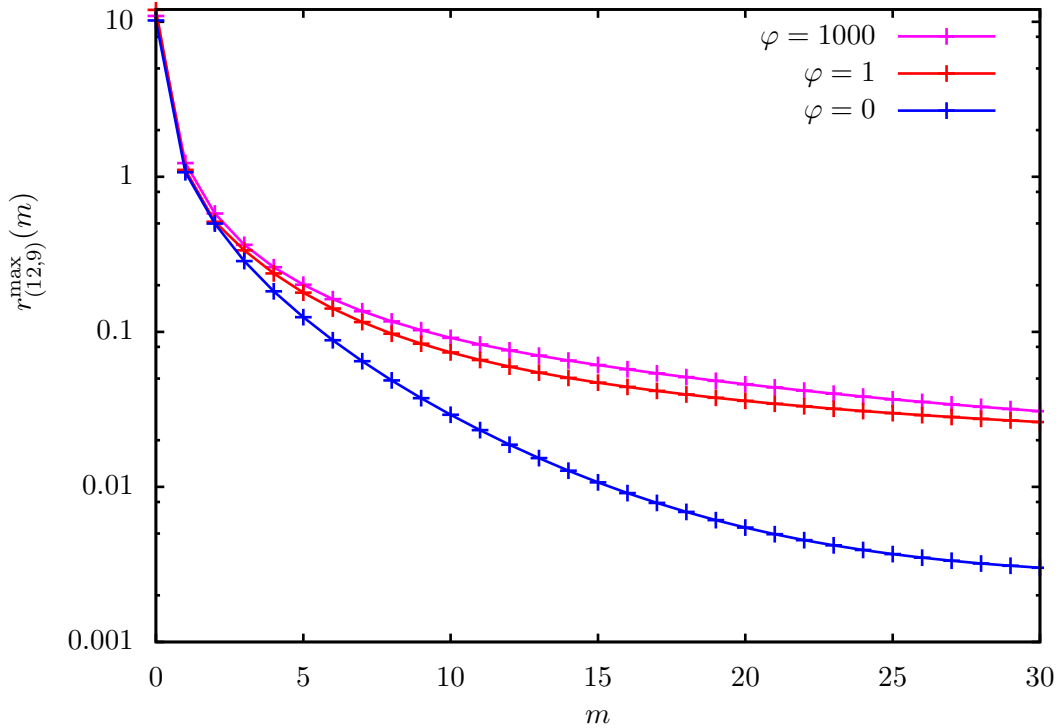}
 % }
  \caption{The convergence/divergence of T-iteration associated with the $SU(2)$ bundle in \eqref{k3_eg1} at the stability wall in K\"ahler moduli space. The bundle modulus $\varphi$ defined in \eqref{f_varphi} determines whether $\Vsheaf$ is strictly semi-stable (when $\varphi \neq 0$) or poly-stable (when $\varphi=0$). In the latter case, the reducible connection satisfies the Hermitian Yang-Mills equations \eqref{the_hym}.    
    }
  \label{tab:r_iterations}
\end{figure}
One can quantify this decomposable locus in the bundle moduli space as
follows. The bundle moduli space of $\Vsheaf$ is described by the
parameters in the map $f$ in \eqref{k3_eg1}. We can describe the
decomposable locus in \eqref{split} in terms of this map by
parametrizing one particular vector bundle modulus as $\varphi$ in the
monad map
\begin{equation}\label{f_varphi}
  f_\varphi 
  =
  \begin{pmatrix}
    (9-i) y_0 + (-6-6i) y_1 + (-8-8i) y_2 \\
    (-9-9i) y_0 + (-2+9i) y_1 + ( 4-4i) y_2 \\
    \varphi 
    \big(
    { \scriptstyle
      (-10+7i)     x_0^4 + 
      (  6-9i)     x_0^3 x_1+
      (  3-8i)     x_0^2 x_1^2+
      ( -9-4i)     x_0 x_1^3+
      (  8+4i)     x_1^4
    }
    \big)
  \end{pmatrix}
\end{equation}
which determines $\Vsheaf_{\varphi}$,
\begin{equation}
  \label{eq:Vphi}
  \vcenter{\xymatrix{
      0
      \ar[r] &
      \Osheaf(-2,-1)
      \ar[r]^-{f_\varphi} &
       \Osheaf(-2,0)^{\oplus 2} + \Osheaf(2,-1)
      \ar[r] &
      \Vsheaf_\varphi
      \ar[r] &
      0~.
    }}
\end{equation}
When $\varphi=0$, the bundle splits as in \eqref{split} and the
resulting direct sum is poly-stable and a solution to the Hermitian
Yang-Mills equations. When $\varphi \neq 0$, $\Vsheaf$ is only
semi-stable, one cannot solve \eqref{the_hym}, and supersymmetry is
broken.

Let us now revisit the numerical results of the past few sections in light of this structure. Can the error measures $\tau(A_{\Vsheaf})$ and $r^{max}(m)$, introduced in \autoref{modific} and \autoref{t_iter_error} respectively, accurately reveal this $\varphi$-dependent structure? To answer this question, we have run the algorithm for the same bundle again, this time keeping the K\"ahler moduli fixed directly on the line $(t_1,t_2) \sim (4,3)$, but allowing the bundle moduli to vary through the variable $\varphi$. The metric was computed for this polarization with $N_g = 202,800$ points (adaptive), $30$ iterations of the metric T-operator and with a K\"ahler form determined by $\cO(4,3)^{\otimes 3}$. Meanwhile, the connection utilized $N_G = 74,892$ points (adaptive) and $100$ iterations of the connection T-operator.
The results for the $\tau(t_1,t_2)$ error measure of equation \eqref{tau_norm} are shown in \autoref{tab:Wall}. Significantly, the algorithm clearly distinguishes between the semi-stable ($\varphi \neq 0$) and poly-stable ($\varphi=0$) behavior of $\Vsheaf$. Moreover, one can repeat the same analysis with the T-operator convergence measure, $r^{max}(12,9)(m)$, of \eqref{r_def}. For this analysis, we hold ourselves fixed at one point in K\"ahler moduli space ($t_1=12, t_2=9$) and allowing the number of T-iterations $m$ to increase. Once again, we find that this error measure definitively distinguishes between the supersymmetric and non-supersymmetric configurations. The convergence of the T-iteration is shown in \autoref{tab:r_iterations}.

The graph \autoref{tab:r_iterations} shows a high degree of
sensitivity to the bundle moduli. The fact that the T-iterations can
distinguish the $\varphi=0$ and $\varphi \neq 0$ cases is significant,
since it means that for a general value of the bundle moduli $\Vsheaf$
is \emph{not} Gieseker stable for the boundary wall polarization. This
can be verified directly using the destabilizing subsheaf $\cF$ in
\eqref{destabilizing} and the definitions of Gieseker stability,
\eqref{gieseker}, in \autoref{review_section}. In fact, for
$\cL=\cO(4,3)$ we have
\begin{equation}
  p_{\cL}(\cF)(n)=45n^2-3  ,
\end{equation}
while 
\begin{equation}
  p_{\cL}(\Vsheaf)(n)=\frac{1}{2}(90n^2-8)~.
\end{equation}
Therefore, 
\begin{equation}
 p_{\cL}(\Vsheaf)(n)\prec p_{\cL}(\cF)(n) 
\end{equation}
and, by \eqref{gieseker_ineq}, $\Vsheaf$ is not Gieseker stable. The results of \autoref{tab:r_iterations} are in complete agreement with this; showing the $\varphi=0$ reducible connection T-operator converging to its fixed point while the Gieseker unstable configurations with $\varphi \neq 0$ fail to converge. It follows that the T-operator is sensitive enough to the bundle moduli that it can distinguish between these important cases.

\section{A Calabi-Yau Threefold Example}
\label{cy3_sec}

In the previous sections, we investigated K\"ahler cone substructure with a number of new tools and from a variety of perspectives. Here, we present a final example involving geometry directly relevant to realistic $\Ncal=1$ heterotic compactifications; namely, a Calabi-Yau threefold with a higher-rank vector bundle defined over it.

As a base manifold, we choose a Calabi-Yau threefold determined by a
generic degree $(4,2)$-hypersurface in $\CP^3\times\CP^1$. For this
threefold, $h^{1,1}=2$ and all line bundles can be written in the form
$\cO(a_1,a_2)$, where $\cO(1,0)$ arises from the hyperplane descending
from $\mathbb{P}^1$ and $\cO(0,1)$ descends from the hyperplane of
$\mathbb{P}^3$.
\begin{figure}[htb]
  \centering
 % \framebox{
  \input{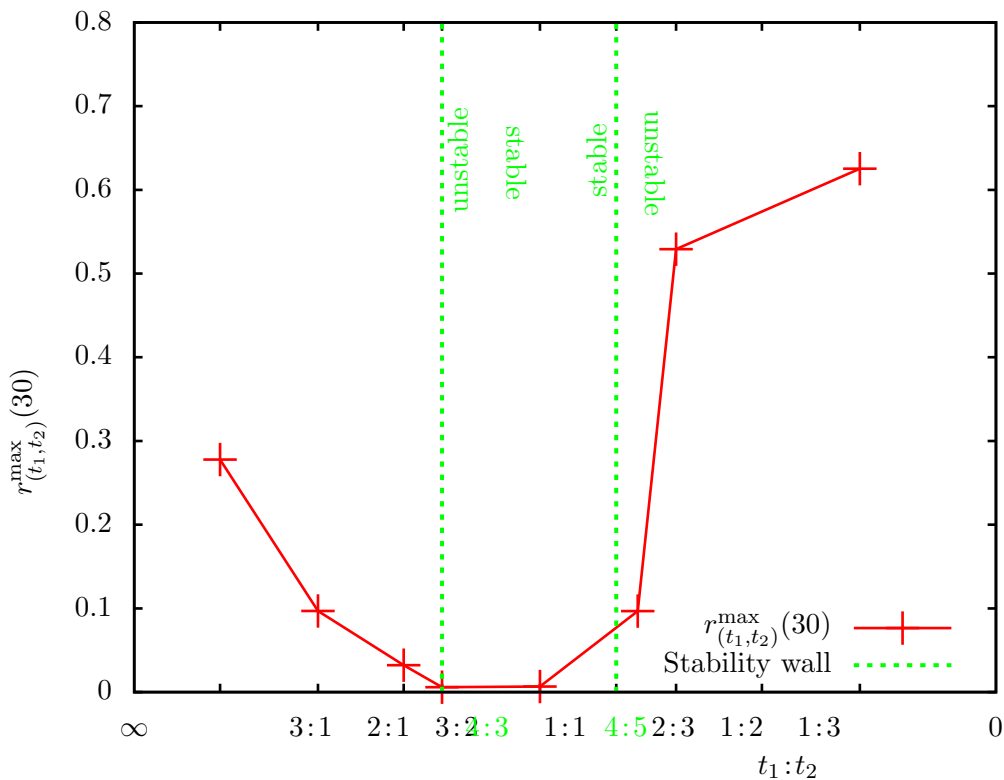}
 % }
  \caption{Convergence of the T-iteration on the $SU(3)$ bundle defined by \eqref{cy3_eg_dual} for different
    K\"ahler moduli. The radial direction $\gcd(t_1,t_2)$ is always
    chosen to be as large as possible subject to the constraint that
    there are $\dim H^0\big( \Osheaf(t_1,t_2) \big) \leq 500$
    sections. The results agree with the K\"ahler cone substructure in \eqref{cy3substruc}.}
  \label{tab:AngularThreefold}
\end{figure}
Over this threefold, define the $SU(3)$ monad bundle
\begin{equation}\label{cy3_eg}
  0  \longrightarrow  \Vsheaf  \longrightarrow  \cO(-2,1) \oplus \cO(3,-1) \oplus \cO(2,0) \oplus \cO(2,1)^{\oplus 2} \stackrel{f}{\longrightarrow} \cO(4,1) \oplus \cO(3,1)  \longrightarrow  0
\end{equation}
and its associated dual bundle
\begin{multline}
  \label{cy3_eg_dual}
  0 \longrightarrow \cO(-4,-1) \oplus
  \cO(-3,-1) \longrightarrow 
  \\
  \cO(-2,-1)^{\oplus 2} \oplus \cO(-2,0)
  \oplus \cO(-3,1) \oplus \cO(2,-1) \longrightarrow \Vsheaf^{\vee}
  \longrightarrow 0 .
\end{multline}
This example was chosen because it once again the stability depends on
the precise value of the K\"ahler moduli. It can be verified, using
the definitions of \autoref{review_section}, that the bundle $\Vsheaf$
in eq.~\eqref{cy3_eg} has two potentially destabilizing
sub-sheaves. These are $\cF_1$,$\cF_2 \subset \Vsheaf$ given by
\begin{equation}
  \begin{split}
    &0 \longrightarrow \cF_1 \longrightarrow 
    \cO(3,-1) \oplus \cO(2,0) \oplus \cO(2,1)^{\oplus 2} 
    \stackrel{f}{\longrightarrow} 
    \cO(4,1) \oplus \cO(3,1) \longrightarrow 0 ,\\
    &0 \longrightarrow \cF_2 \longrightarrow 
    \cO(-2,1) \oplus \cO(2,0) \oplus \cO(2,1)^{\oplus 2} 
    \stackrel{f}{\longrightarrow} \cO(4,1)
    \oplus \cO(3,1) \longrightarrow 0
    .
  \end{split}
\end{equation}
Both are of rank $2$ with $c_1(\cF_1)=(2,-1)$ and
$c_1(\cF_2)=(-3,1)$. Due to these two destabilizing sub-sheaves, the
K\"ahler cone divides into three regions
\begin{equation}\label{cy3substruc}
  \begin{split}
    \mathcal{K}_\text{stable} 
    =&\;
    \big\{
    \Osheaf(t_1, t_2)
    \;\big|\;
    \tfrac{4}{5} < \tfrac{t_2}{t_1} < \tfrac{4}{3}
    \big\}
    , \\
    \mathcal{K}_\text{unstable} 
    =&\;
    \big\{
    \Osheaf(t_1, t_2)
    \;\big|\;
    \tfrac{t_2}{t_1} < \tfrac{4}{5}
    \big\}
    ~\cup~
    \big\{
    \Osheaf(t_1, t_2)
    \;\big|\;
    \tfrac{t_2}{t_1} > \tfrac{4}{3}
    \big\}  
    .
  \end{split}
\end{equation}
Moreover, a direct calculation using \eqref{gieseker} shows that
$\Vsheaf$ is \emph{still} Gieseker stable at each of the stability
walls defined by $t_2/t_1=4/3$ and $t_2/t_1=4/5$. Thus, while the
bundle is strictly semi-stable along these rays (and, hence, does not
solve the Hermitian Yang-Mills equations eq.~\eqref{the_hym}), one would
still expect the T-iteration to converge for points along these
boundaries.

For ease of embedding, the T-operator is computed for
$\Vsheaf^{\vee}$. However, since $\Vsheaf$ and its dual must be
slope-stable/unstable in the same regions of K\"ahler moduli space,
this does not affect our results.  Since we have demonstrated in the
previous sections that the T-iteration is a fast and efficient check
of stability, we will consider here only the error measure
$r_{(t_1,t_2)}^{max}(m)$. It is interesting to observe that, even in
this more complex example, the data shown in
\autoref{tab:AngularThreefold} clearly reproduces the K\"ahler cone
substructure described in \eqref{cy3substruc}. Moreover, at the points
on the two stability walls we find, as expected, that
$r_{(t_1,t_2)}^{max}(30)=0$, since $\Vsheaf$ is still Gieseker stable
for these lines of slope semi-stability. The connection integration
was performed with $N_G = 119,164$ points (adaptive) and $30$
iterations of the T-operator. The plot shown in
\autoref{tab:AngularThreefold} are analogous to the results in
\autoref{tab:Angular} for the $K3$ surface.

\section{Conclusions and Future Work}\label{conclusion_sec}

In this paper, we extended the generalized Donaldson algorithm to manifolds with 
higher-dimensional K\"ahler cones and presented a method for approximating the connection on slope-stable holomorphic vector bundles in this context. We also introduced a numerical criterion for determining the existence of supersymmetric heterotic vacua, without having to compute the connection.
These techniques clearly can be used to search for new classes of smooth $\Ncal=1$ supersymmetric  vacua in heterotic string theory. However, the explicit knowledge of the gauge connection that they provide allows us to go far beyond a simple categorization of vacua.

It is a long-standing goal of string theory to produce low-energy
theories whose effective actions reproduce the symmetries, particle
spectrum, and properties of elementary particle physics. Within the
context of smooth heterotic string or M-theory
compactifications~\cite{Candelas:1985en, Lukas:1998yy, Lukas:1998ew},
there has been considerable progress in recent
years~\cite{Braun:2004xv, Braun:2005nv, Braun:2005ux, Braun:2005zv,
  Buchbinder:2007ad, Bouchard:2005ag, Anderson:2009mh}.  Despite these
successes, certain observable quantities of particle physics, such as
the gauge and Yukawa couplings, are difficult to compute
directly. Normalized Yukawa couplings, for example, depend on both the
coefficients of cubic terms in the superpotential and the explicit
form of the K\"ahler potential. In turn, these quantities depend on
the detailed structure of the underlying geometry -- that is, the
metric and gauge connection on the Calabi-Yau threefold and
holomorphic vector bundle respectively. Hence, the Yukawa couplings in
the four-dimensional effective theory are not known, except in very
special cases. One can rarely do better than the qualitative statement
that such coupling coefficents either vanish or are of order
one. However, the methods introduced
in~\cite{Douglas:2006rr,Anderson:2010ke} and extended in this paper
allow one, for the first time, to explicitly compute both the metric
and the gauge connection and, hence, the Yukawa coefficients. In
future work~\cite{yukawas_numeric}, we will use these techniques to
calculate these physical couplings.

\section*{Acknowledgments}

The work of L. Anderson and B. A. Ovrut is supported in part by the DOE under contract No. DE-AC02-76-ER-$03071$ and by NSF RTG Grant DMS-$0636606$ and NSF-$1001296$.

\bibliographystyle{utcaps} %utcaps....%
\renewcommand{\refname}{Bibliography}
\addcontentsline{toc}{section}{Bibliography} 

\bibliography{Volker,Metric,Part}

\providecommand{\href}[2]{#2}\begingroup\raggedright\begin{thebibliography}{10}

\bibitem{Gross:1984dd}
D.~J. Gross, J.~A. Harvey, E.~J. Martinec, and R.~Rohm, ``The Heterotic
  String,'' {\em Phys. Rev. Lett.} {\bf 54} (1985)
502--505.
%%CITATION = PRLTA,54,502;%%.

\bibitem{Candelas:1985en}
P.~Candelas, G.~T. Horowitz, A.~Strominger, and E.~Witten, ``Vacuum
  Configurations for Superstrings,'' {\em Nucl. Phys.} {\bf B258} (1985)
46--74.
%%CITATION = NUPHA,B258,46;%%.

\bibitem{Horava:1995qa}
P.~Horava and E.~Witten, ``Heterotic and type I string dynamics from eleven
  dimensions,'' {\em Nucl. Phys.} {\bf B460} (1996) 506--524,
\href{http://arXiv.org/abs/hep-th/9510209}{{\tt hep-th/9510209}}.
%%CITATION = HEP-TH 9510209;%%.

\bibitem{Horava:1996ma}
P.~Horava and E.~Witten, ``Eleven-Dimensional Supergravity on a Manifold with
  Boundary,'' {\em Nucl. Phys.} {\bf B475} (1996) 94--114,
\href{http://arXiv.org/abs/hep-th/9603142}{{\tt hep-th/9603142}}.
%%CITATION = HEP-TH 9603142;%%.

\bibitem{Witten:1996mz}
E.~Witten, ``Strong Coupling Expansion Of Calabi-Yau Compactification,'' {\em
  Nucl. Phys.} {\bf B471} (1996) 135--158,
\href{http://arXiv.org/abs/hep-th/9602070}{{\tt hep-th/9602070}}.
%%CITATION = HEP-TH 9602070;%%.

\bibitem{Lukas:1997fg}
A.~Lukas, B.~A. Ovrut, and D.~Waldram, ``On the four-dimensional effective
  action of strongly coupled heterotic string theory,'' {\em Nucl. Phys.} {\bf
  B532} (1998) 43--82,
\href{http://arXiv.org/abs/hep-th/9710208}{{\tt hep-th/9710208}}.
%%CITATION = HEP-TH 9710208;%%.

\bibitem{Lukas:1998yy}
A.~Lukas, B.~A. Ovrut, K.~S. Stelle, and D.~Waldram, ``{The universe as a
  domain wall},'' {\em Phys. Rev.} {\bf D59} (1999) 086001,
\href{http://arXiv.org/abs/hep-th/9803235}{{\tt hep-th/9803235}}.
%%CITATION = HEP-TH/9803235;%%.

\bibitem{Lukas:1998tt}
A.~Lukas, B.~A. Ovrut, K.~S. Stelle, and D.~Waldram, ``Heterotic M-theory in
  five dimensions,'' {\em Nucl. Phys.} {\bf B552} (1999) 246--290,
\href{http://arXiv.org/abs/hep-th/9806051}{{\tt hep-th/9806051}}.
%%CITATION = HEP-TH 9806051;%%.

\bibitem{Green:1987sp}
M.~B. Green, J.~H. Schwarz, and E.~Witten, ``SUPERSTRING THEORY. VOL. 1:
  INTRODUCTION,''. Cambridge, Uk: Univ. Pr. (1987) 469 P. (Cambridge Monographs
  On Mathematical Physics).

\bibitem{MR480350}
S.~T. Yau, ``On the {R}icci curvature of a compact {K}\"ahler manifold and the
  complex {M}onge-{A}mp\`ere equation. {I},'' {\em Comm. Pure Appl. Math.} {\bf
  31} (1978), no.~3, 339--411.

\bibitem{duy1}
K.~Uhlenbeck and S.-T. Yau., ``{On the existence of Hermitian Yang-Mills
  connections in stable bundles},'' {\em Comm. Pure App. Math.} {\bf 39} (1986)
  257.

\bibitem{duy2}
S.~Donaldson, ``{Anti Self-Dual Yang-Mills Connections over Complex Algebraic
  Surfaces and Stable Vector Bundles,},'' {\em Proc. London Math. Soc.} {\bf 3}
  (1985) 1.

\bibitem{Candelas:1987rx}
P.~Candelas and S.~Kalara, ``Yukawa couplings for a three generation
  superstring compactification,'' {\em Nucl. Phys.} {\bf B298} (1988)
357.
%%CITATION = NUPHA,B298,357;%%.

\bibitem{Candelas:1990rm}
P.~Candelas, X.~C. De~La~Ossa, P.~S. Green, and L.~Parkes, ``A pair of
  Calabi-Yau manifolds as an exactly soluble superconformal theory,'' {\em
  Nucl. Phys.} {\bf B359} (1991)
21--74.
%%CITATION = NUPHA,B359,21;%%.

\bibitem{Greene:1993vm}
B.~R. Greene, D.~R. Morrison, and M.~R. Plesser, ``Mirror manifolds in higher
  dimension,'' {\em Commun. Math. Phys.} {\bf 173} (1995) 559--598,
\href{http://arXiv.org/abs/hep-th/9402119}{{\tt hep-th/9402119}}.
%%CITATION = HEP-TH/9402119;%%.

\bibitem{Donagi:2006yf}
R.~Donagi, R.~Reinbacher, and S.-T. Yau, ``Yukawa couplings on quintic
  threefolds,''
\href{http://arXiv.org/abs/hep-th/0605203}{{\tt hep-th/0605203}}.
%%CITATION = HEP-TH/0605203;%%.

\bibitem{Braun:2006me}
V.~Braun, Y.-H. He, and B.~A. Ovrut, ``Yukawa couplings in heterotic standard
  models,'' {\em JHEP} {\bf 04} (2006) 019,
\href{http://arXiv.org/abs/hep-th/0601204}{{\tt hep-th/0601204}}.
%%CITATION = HEP-TH/0601204;%%.

\bibitem{Anderson:2009ge}
L.~B. Anderson, J.~Gray, D.~Grayson, Y.-H. He, and A.~Lukas, ``{Yukawa
  Couplings in Heterotic Compactification},'' {\em Commun. Math. Phys.} {\bf
  297} (2010) 95--127,
\href{http://arXiv.org/abs/0904.2186}{{\tt 0904.2186}}.
%%CITATION = 0904.2186;%%.

\bibitem{MR2161248}
S.~K. Donaldson, ``Scalar curvature and projective embeddings. {II},'' {\em Q.
  J. Math.} {\bf 56} (2005), no.~3, 345--356.

\bibitem{MR1916953}
S.~K. Donaldson, ``Scalar curvature and projective embeddings. {I},'' {\em J.
  Differential Geom.} {\bf 59} (2001), no.~3, 479--522.

\bibitem{DonaldsonNumerical}
S.~K. Donaldson, ``Some numerical results in complex differential geometry,''
  \href{http://arXiv.org/abs/math.DG/0512625}{{\tt math.DG/0512625}}.

\bibitem{MR1064867}
G.~Tian, ``On a set of polarized {K}\"ahler metrics on algebraic manifolds,''
  {\em J. Differential Geom.} {\bf 32} (1990), no.~1, 99--130.

\bibitem{MR2154820}
X.~Wang, ``Canonical metrics on stable vector bundles,'' {\em Comm. Anal.
  Geom.} {\bf 13} (2005), no.~2, 253--285.

\bibitem{Douglas:2006hz}
M.~R. Douglas, R.~L. Karp, S.~Lukic, and R.~Reinbacher, ``Numerical solution to
  the hermitian Yang-Mills equation on the Fermat quintic,''
\href{http://arXiv.org/abs/hep-th/0606261}{{\tt hep-th/0606261}}.
%%CITATION = HEP-TH/0606261;%%.

\bibitem{MR2194329}
M.~Headrick and T.~Wiseman, ``Numerical {R}icci-flat metrics on {$K3$},'' {\em
  Classical Quantum Gravity} {\bf 22} (2005), no.~23, 4931--4960.

\bibitem{Doran:2007zn}
C.~Doran, M.~Headrick, C.~P. Herzog, J.~Kantor, and T.~Wiseman, ``Numerical
  Kaehler-Einstein metric on the third del Pezzo,''
\href{http://arXiv.org/abs/hep-th/0703057}{{\tt hep-th/0703057}}.
%%CITATION = HEP-TH/0703057;%%.

\bibitem{Headrick:2009jz}
M.~Headrick and A.~Nassar, ``{Energy functionals for Calabi-Yau metrics},''
  \href{http://arXiv.org/abs/0908.2635}{{\tt 0908.2635}}.

\bibitem{Douglas:2008es}
M.~R. Douglas and S.~Klevtsov, ``{Black holes and balanced metrics},''
\href{http://arXiv.org/abs/0811.0367}{{\tt 0811.0367}}.
%%CITATION = 0811.0367;%%.

\bibitem{Anderson:2010ke}
L.~B. Anderson, V.~Braun, R.~L. Karp, and B.~A. Ovrut, ``{Numerical Hermitian
  Yang-Mills Connections and Vector Bundle Stability in Heterotic Theories},''
  {\em JHEP} {\bf 06} (2010) 107,
\href{http://arXiv.org/abs/1004.4399}{{\tt 1004.4399}}.
%%CITATION = 1004.4399;%%.

\bibitem{Braun:2007sn}
V.~Braun, T.~Brelidze, M.~R. Douglas, and B.~A. Ovrut, ``Calabi-Yau Metrics for
  Quotients and Complete Intersections,''
\href{http://arXiv.org/abs/arXiv:0712.3563 [hep-th]}{{\tt arXiv:0712.3563
  [hep-th]}}.
%%CITATION = ARXIV:0712.3563;%%.

\bibitem{Braun:2008jp}
V.~Braun, T.~Brelidze, M.~R. Douglas, and B.~A. Ovrut, ``{Eigenvalues and
  Eigenfunctions of the Scalar Laplace Operator on Calabi-Yau Manifolds},''
  {\em JHEP} {\bf 07} (2008) 120,
\href{http://arXiv.org/abs/0805.3689}{{\tt 0805.3689}}.
%%CITATION = 0805.3689;%%.

\bibitem{Douglas:2006rr}
M.~R. Douglas, R.~L. Karp, S.~Lukic, and R.~Reinbacher, ``Numerical Calabi-Yau
  metrics,''
\href{http://arXiv.org/abs/hep-th/0612075}{{\tt hep-th/0612075}}.
%%CITATION = HEP-TH/0612075;%%.

\bibitem{2008arXiv0804.4005S}
R.~{Seyyedali}, ``{Numerical Algorithms for Finding Balanced Metrics on Vector
  Bundles},'' {\em ArXiv e-prints} (Apr., 2008)
  \href{http://arXiv.org/abs/0804.4005}{{\tt 0804.4005}}.

\bibitem{2007arXiv0709.1490K}
J.~{Keller}, ``{Ricci iterations on Kahler classes},'' {\em ArXiv e-prints}
  (Sept., 2007) \href{http://arXiv.org/abs/0709.1490}{{\tt 0709.1490}}.

\bibitem{2002math......3254P}
D.~H. {Phong} and J.~{Sturm}, ``{Stability, energy functionals, and
  K$\backslash$''ahler-Einstein metrics},'' {\em ArXiv Mathematics e-prints}
  (Mar., 2002) \href{http://arXiv.org/abs/arXiv:math/0203254}{{\tt
  arXiv:math/0203254}}.

\bibitem{okonek}
H.~S. C.~Okonek, M.~Schneider, {\em {Vector Bundles on Complex Projective
  Spaces}}.
\newblock Birkhauser Verlag, 1988.

\bibitem{Anderson:2007nc}
L.~B. Anderson, Y.-H. He, and A.~Lukas, ``{Heterotic compactification, an
  algorithmic approach},'' {\em JHEP} {\bf 07} (2007) 049,
\href{http://arXiv.org/abs/hep-th/0702210}{{\tt hep-th/0702210}}.
%%CITATION = HEP-TH/0702210;%%.

\bibitem{Anderson:2008uw}
L.~B. Anderson, Y.-H. He, and A.~Lukas, ``{Monad Bundles in Heterotic String
  Compactifications},'' {\em JHEP} {\bf 07} (2008) 104,
\href{http://arXiv.org/abs/0805.2875}{{\tt 0805.2875}}.
%%CITATION = 0805.2875;%%.

\bibitem{Anderson:2009sw}
L.~B. Anderson, J.~Gray, A.~Lukas, and B.~Ovrut, ``{The Edge Of Supersymmetry:
  Stability Walls in Heterotic Theory},'' {\em Phys. Lett.} {\bf B677} (2009)
  190--194,
\href{http://arXiv.org/abs/0903.5088}{{\tt 0903.5088}}.
%%CITATION = 0903.5088;%%.

\bibitem{Anderson:2009nt}
L.~B. Anderson, J.~Gray, A.~Lukas, and B.~Ovrut, ``{Stability Walls in
  Heterotic Theories},'' {\em JHEP} {\bf 09} (2009) 026,
\href{http://arXiv.org/abs/0905.1748}{{\tt 0905.1748}}.
%%CITATION = 0905.1748;%%.

\bibitem{Blumenhagen:2006ux}
R.~Blumenhagen, S.~Moster, and T.~Weigand, ``Heterotic GUT and Standard Model
  vacua from simply connected Calabi-Yau manifolds,''
\href{http://arXiv.org/abs/hep-th/0603015}{{\tt hep-th/0603015}}.
%%CITATION = HEP-TH 0603015;%%.

\bibitem{huybrechts}
D.~Huybrechts and M.~Lehn, ``{The geometry of the Moduli Spaces of Sheaves},''
  {\em Aspects of Mathematics} {\bf E 31} (1997).

\bibitem{MR2283416}
Y.~Sano, ``Numerical algorithm for finding balanced metrics,'' {\em Osaka J.
  Math.} {\bf 43} (2006), no.~3, 679--688.

\bibitem{MR507725}
P.~Griffiths and J.~Harris, {\em Principles of algebraic geometry}.
\newblock Wiley-Interscience [John Wiley \& Sons], New York, 1978.
\newblock Pure and Applied Mathematics.

\bibitem{MR2180559}
J.~Keller, ``Canonical metrics and {H}arder-{N}arasimhan filtration,'' in {\em
  Contemporary aspects of complex analysis, differential geometry and
  mathematical physics}, pp.~132--148.
\newblock World Sci. Publ., Hackensack, NJ, 2005.

\bibitem{Sharpe:1998zu}
E.~R. Sharpe, ``{Kaehler cone substructure},'' {\em Adv. Theor. Math. Phys.}
  {\bf 2} (1999) 1441--1462,
\href{http://arXiv.org/abs/hep-th/9810064}{{\tt hep-th/9810064}}.
%%CITATION = HEP-TH/9810064;%%.

\bibitem{Anderson:2010tc}
L.~B. Anderson, J.~Gray, and B.~Ovrut, ``{Yukawa Textures From Heterotic
  Stability Walls},'' {\em JHEP} {\bf 05} (2010) 086,
\href{http://arXiv.org/abs/1001.2317}{{\tt 1001.2317}}.
%%CITATION = 1001.2317;%%.

\bibitem{Anderson:2010mh}
L.~B. Anderson, J.~Gray, A.~Lukas, and B.~Ovrut, ``{Stabilizing the Complex
  Structure in Heterotic Calabi-Yau Vacua},'' {\em JHEP} {\bf 02} (2011) 088,
\href{http://arXiv.org/abs/1010.0255}{{\tt 1010.0255}}.
%%CITATION = 1010.0255;%%.

\bibitem{Anderson:2011cz}
L.~B. Anderson, J.~Gray, A.~Lukas, and B.~Ovrut, ``{Stabilizing All Geometric
  Moduli in Heterotic Calabi-Yau Vacua},''
\href{http://arXiv.org/abs/1102.0011}{{\tt 1102.0011}}.
%%CITATION = 1102.0011;%%.

\bibitem{Thomas:math9806111}
R.~P. Thomas, ``A holomorphic Casson invariant for Calabi-Yau 3-folds, and
  bundles on K3 fibrations,'' \href{http://arXiv.org/abs/math/9806111}{{\tt
  math/9806111}}.

\bibitem{2010arXiv1002.4080L}
W.~{Li} and Z.~{Qin}, ``{Donaldson-Thomas invariants of certain Calabi-Yau
  3-folds},'' {\em ArXiv e-prints} (Feb., 2010)
  \href{http://arXiv.org/abs/1002.4080}{{\tt 1002.4080}}.

\bibitem{Anderson:2010ty}
L.~B. Anderson, J.~Gray, and B.~Ovrut, ``{Transitions in the Web of Heterotic
  Vacua},''
\href{http://arXiv.org/abs/1012.3179}{{\tt 1012.3179}}.
%%CITATION = 1012.3179;%%.

\bibitem{Anderson:2009mh}
L.~B. Anderson, J.~Gray, Y.-H. He, and A.~Lukas, ``{Exploring Positive Monad
  Bundles And A New Heterotic Standard Model},'' {\em JHEP} {\bf 02} (2010)
  054,
\href{http://arXiv.org/abs/0911.1569}{{\tt 0911.1569}}.
%%CITATION = 0911.1569;%%.

\bibitem{Anderson:2008ex}
L.~B. Anderson, ``{Heterotic and M-theory Compactifications for String
  Phenomenology},''
\href{http://arXiv.org/abs/0808.3621}{{\tt 0808.3621}}.
%%CITATION = 0808.3621;%%.

\bibitem{2005math.....12625D}
S.~K. {Donaldson}, ``{Some numerical results in complex differential
  geometry},'' {\em ArXiv Mathematics e-prints} (Dec., 2005)
  \href{http://arXiv.org/abs/arXiv:math/0512625}{{\tt arXiv:math/0512625}}.

\bibitem{2008arXiv0803.0987B}
R.~S. {Bunch} and S.~K. {Donaldson}, ``{Numerical approximations to extremal
  metrics on toric surfaces},'' {\em ArXiv e-prints} (Mar., 2008)
  \href{http://arXiv.org/abs/0803.0987}{{\tt 0803.0987}}.

\bibitem{Zelditch:Shif}
B.~Shiffman and S.~Zelditch, ``Distribution of zeros of random and quantum
  chaotic sections of positive line bundles,'' {\em Comm. Math. Phys.} {\bf
  200} (1999), no.~3, 661--683.

\bibitem{MR1616718}
S.~Zelditch, ``Szeg{\H o} kernels and a theorem of {T}ian,'' {\em Internat.
  Math. Res. Notices} (1998), no.~6, 317--331.

\bibitem{Keller:2009vj}
J.~Keller and S.~Lukic, ``{Numerical Weil-Petersson metrics on moduli spaces of
  Calabi-Yau manifolds},'' \href{http://arXiv.org/abs/0907.1387}{{\tt
  0907.1387}}.

\bibitem{Lukas:1998ew}
A.~Lukas, B.~A. Ovrut, and D.~Waldram, ``{The ten-dimensional effective action
  of strongly coupled heterotic string theory},'' {\em Nucl. Phys.} {\bf B540}
  (1999) 230--246,
\href{http://arXiv.org/abs/hep-th/9801087}{{\tt hep-th/9801087}}.
%%CITATION = HEP-TH/9801087;%%.

\bibitem{Braun:2004xv}
V.~Braun, B.~A. Ovrut, T.~Pantev, and R.~Reinbacher, ``Elliptic Calabi-Yau
  threefolds with Z(3) x Z(3) Wilson lines,'' {\em JHEP} {\bf 12} (2004) 062,
\href{http://arXiv.org/abs/hep-th/0410055}{{\tt hep-th/0410055}}.
%%CITATION = HEP-TH/0410055;%%.

\bibitem{Braun:2005nv}
V.~Braun, Y.-H. He, B.~A. Ovrut, and T.~Pantev, ``The exact MSSM spectrum from
  string theory,'' {\em JHEP} {\bf 05} (2006) 043,
\href{http://arXiv.org/abs/hep-th/0512177}{{\tt hep-th/0512177}}.
%%CITATION = HEP-TH/0512177;%%.

\bibitem{Braun:2005ux}
V.~Braun, Y.-H. He, B.~A. Ovrut, and T.~Pantev, ``A heterotic standard model,''
  {\em Phys. Lett.} {\bf B618} (2005) 252--258,
\href{http://arXiv.org/abs/hep-th/0501070}{{\tt hep-th/0501070}}.
%%CITATION = HEP-TH/0501070;%%.

\bibitem{Braun:2005zv}
V.~Braun, Y.-H. He, B.~A. Ovrut, and T.~Pantev, ``Vector bundle extensions,
  sheaf cohomology, and the heterotic standard model,'' {\em Adv. Theor. Math.
  Phys.} {\bf 10} (2006) 4,
\href{http://arXiv.org/abs/hep-th/0505041}{{\tt hep-th/0505041}}.
%%CITATION = HEP-TH/0505041;%%.

\bibitem{Buchbinder:2007ad}
E.~I. Buchbinder, J.~Khoury, and B.~A. Ovrut, ``{New Ekpyrotic Cosmology},''
  {\em Phys. Rev.} {\bf D76} (2007) 123503,
\href{http://arXiv.org/abs/hep-th/0702154}{{\tt hep-th/0702154}}.
%%CITATION = HEP-TH/0702154;%%.

\bibitem{Bouchard:2005ag}
V.~Bouchard and R.~Donagi, ``An SU(5) heterotic standard model,'' {\em Phys.
  Lett.} {\bf B633} (2006) 783--791,
\href{http://arXiv.org/abs/hep-th/0512149}{{\tt hep-th/0512149}}.
%%CITATION = HEP-TH/0512149;%%.

\bibitem{yukawas_numeric}
L.~B. Anderson, V.~Braun, and B.~A. Ovrut, ``{Numerical determination of Yukawa
  couplings}.'' {To appear.}

\end{thebibliography}\endgroup

\end{document}